\documentclass[twocolumn,tighten,times]{aastex631}
\usepackage[T1]{fontenc}
\usepackage{pgffor}
\usepackage{tablefootnote} 
\usepackage{soul}

\newcommand\ergs{erg~s$^{-1}$}
\newcommand\ergcms{erg~cm$^{-2}$~s$^{-1}$}
\newcommand\mjybeam{mJy~beam$^{-1}$}
\newcommand{\HII}{H\,{\sc ii}}
\newcommand{\FeII}{[Fe\,{\sc ii}]\,$\lambda$16440}

\newcommand\XOR{$\log(f_{\rm X}/f_V)$}
\newcommand{\hbeta}{H{$\beta$}}

\shorttitle{Radio counterparts of ULXs}
\shortauthors{Zhang et al.}

\graphicspath{{./}{fig/}}

\begin{document}

\title{Radio counterparts of ultraluminous X-ray sources in nearby galaxies}

\author[0009-0008-8549-8069]{Yijia Zhang}
\affiliation{Department of Astronomy, Tsinghua University, Beijing 100084, People's Republic of China}

\correspondingauthor{Hua Feng}
\email{hfeng@ihep.ac.cn}

\author[0000-0001-7584-6236]{Hua Feng}
\affiliation{State Key Laboratory of Particle Astrophysics, Institute of High Energy Physics, Chinese Academy of Sciences, Beijing 100049, People's Republic of China}

\author[0000-0002-7351-5801]{Ailing Wang}
\affiliation{State Key Laboratory of Particle Astrophysics, Institute of High Energy Physics, Chinese Academy of Sciences, Beijing 100049, People's Republic of China}

\author[0000-0002-4622-796X]{Roberto Soria}
\affiliation{INAF-Osservatorio Astrofisico di Torino, Pino Torinese, Italy}
%\affiliation{Sydney Institute for Astronomy, School of Physics A28, The University of Sydney, Sydney, NSW, Australia}

\begin{abstract}

Radio surveys of ultraluminous X-ray sources (ULXs) allow us to find supercritically accreting compact objects (SS 433/W50 like systems) or stripped nuclear black holes in nearby galaxies. 
We identified 21 such objects by crossmatching a ULX catalog with the Rapid ASKAP Continuum Survey (RACS) and Very Large Array Sky Survey (VLASS). 
They may have a diverse population. 
(i) Three have a double lobed radio structure with a compact core found in two of them, and could be quasars.
(ii) Five are associated with extended radio structure and star forming regions in optical, where the radio emission is likely due to star forming activities, although the steep radio spectrum up to several GHz casts doubt on that. 
Two of them show X-ray variability suggesting that they are ULXs embedded in star forming regions.
(iii) Thirteen are associated with an unresolved radio source, with a steep spectrum seen in eight, a flat or inverted spectrum seen in two. 
Those with a steep spectrum are arguably candidates for SS 433/W50 like objects, with radio emission due to optically thin synchrotron radiation in a surrounding jet/wind powered nebula. 
Remarkable cases include NGC 925 ULX1 and NGC 6946 ULX1, which are associated with an optical nebula. 
Those with a flat or inverted spectrum could be accreting black holes with a compact jet, while the black hole mass is estimated to be several $10^6 - 10^8$ $M_\sun$ based on the fundamental plane. 
Redshift measurements are needed to firmly determine the association with their apparent host galaxy. 
\end{abstract}

\section{Introduction}

Ultraluminous X-ray sources \citep[ULXs;][]{Kaaret2017, Fabrika2021, King2023} are non-nuclear X-ray sources with an isotropic luminosity exceeding $10^{39}$~\ergs\ or the Eddington limit of a stellar mass black hole. 
The nature of ULXs is still unclear, and they may contain a diverse population \citep{Feng2005, Earnshaw2019, Tranin2024}.

The majority of ULXs are believed to be powered by supercritical accretion onto stellar mass compact objects, as evidenced by the identification of pulsars in ULXs \citep[e.g.][]{Bachetti2014}.
In the regime of supercritical accretion, radiation pressure will be dominant in the accretion flow and drive massive winds \citep{Meier1982, Poutanen2007, Ohsuga2011, Jiang2014, sadowski2016, Kitaki2021}.
X-ray spectroscopy has revealed blueshifted absorption features in the spectrum of ULXs \citep{Walton2016, Pinto2016, Pinto2017, Pinto2021,  Kosec2018}, suggesting that relativistic winds are indeed present in them \citep[for a review see][]{Pinto2023}. 
These outflows will eventually interact with the ambient medium and produce shock-ionized nebulae, seen as optical emission line bubbles surrounding the ULXs \citep{Pakull2002, Pakull2003, Ramsey2006, Abolmasov2007,  Abolmasov2008, Russell2011, Soria2021, Gurpide2022, Zhou2022, Zhou2023a, Zhou2023b}. 

Alternatively, a small fraction of ULXs may contain intermediate mass black holes (IMBHs) with a mass in the range of $10^2 - 10^5$~$M_\sun$ \citep{Colbert1999}.  
In particular, the search of IMBHs focuses on ULXs at the high luminosity end \citep{Sutton2012, Barrows2019, MacKenzie2023}, known as hyper-luminous X-ray sources (HLXs) with a luminosity in excess of $10^{41}$~\ergs\ \citep{Gao2003, Amato2025}.
Remarkable candidates for IMBHs include ESO 243-49 HLX-1 \citep{Farrell2009, Servillat2011} and M82 X-1 \citep{Feng2010, Pasham2014}.

In general, supermassive black holes are not expected at nonnuclear positions due to dynamical friction \citep{Tremaine1975}. 
However, they may reside in ultracompact dwarf galaxies as a result of tidal stripping during galaxy merger, with a black hole mass in the range from IMBH up to $10^8$~$M_\sun$ \citep{Seth2014, Graham2025}.
If they undergo active accretion, they may appear as ULXs. 
However, no such cases have been observed. 

Radio observations of ULXs have produced useful constraints on their nature. 
In most cases, the radio emission is found to be optically thin and extended, spatially coincident with the optical nebula, suggesting that the emission is also produced by shocks due to wind/jet interaction with the interstellar medium \citep{VanDyk1994, Lacey1997, Miller2005, Lang2007, Cseh2012, Urquhart2018, Berghea2020, Soria2021}.
% VanDyk1994, Lacey1997, % NGC 6946
% Miller2005, % Ho II
% Lang2007, % NGC 5408
% Cseh2012, % IC 342
% Urquhart2018, % M51
% Berghea2020, % Ho IX
% Soria2021, % NGC 5585
In a few cases, a compact radio core is detected with the very long baseline interferometry (VLBI), allowing for black hole mass estimation assuming jet emission in the fundamental plane \citep{Mezcua2011, Mezcua2013}.
If this assumption is valid, their typical X-ray and radio luminosities would generally suggest the presence of IMBHs \citep[cf., Table 3 in][]{Mezcua2013}. 
Transient radio emission was detected in HLX-1, offering us the opportunity to constrain the black hole mass to be $< 2.8 \times 10^6$~$M_\sun$ assuming the source was observed in the fundamental plane and considering that the radio emission might have been Doppler boosted \citep{Webb2012, Cseh2015}. 

SS~433, a microquasar in the Milky Way, has long been considered analogous to ULXs, as it is thought to be powered by supercritical accretion from an evolved massive star onto a black hole candidate \citep[for a review see][]{Fabrika2004}. However, the X-ray emission from the compact object is rather faint; it is speculated that thick outflows may have obscured the intrinsic X-ray emission \citep{Fabrika2021}.
% The faintness of X-ray emission in SS~433 could be a result of \red{the} large viewing angle, such that the X-ray flux to the line of sight is obscured by the thick winds \citep{Fabrika2021}. 
SS~433 itself is a bright radio source ($\sim$1~Jy) due to a pair of precessing, relativistic jets, which drive a radio nebula W50 (71~Jy at 1.4 GHz with an extent of about 2\arcdeg) further out when interacting with the surrounding medium \citep{Dubner1998}.
NGC 7793 S26 is one of the best extragalactic analogs to SS~433 and shows a pair of lobes with emission in X-ray, optical emission line, and radio \citep{Pannuti2002, Pakull2010, Soria2010}, driven by shocks with a total mechanical power of about $10^{40}$~\ergs.
Other similar cases include NGC 300 S10 \citep{McLeod2019, Urquhart2019} and M83 MQ1 \citep{Soria2014} and S2 \citep{Soria2020}. 
If SS~433/W50 were located in a nearby galaxy at a distance that is $10^3$ times greater, and we observed the system along its symmetric axis, we would observe a ULX associated with a radio nebula.

Thus, radio observations provide a distinctive means to explore the physics of supercritical accretion or to identify IMBHs.
Previously, radio emission in ULXs have been systematically searched with the Faint Images of the Radio Sky (FIRST) survey~\citep{Sanchez-Sutil2006, Perez-Ramirez2010}, resulting in useful candidates for in-depth case studies. 
The FIRST survey was conducted with VLA at 20~cm with a root-mean-square (rms) of 0.15 mJy and an angular resolution of 5\arcsec. 
Today, radio surveys with a better sensitivity and angular resolution are available, including the Rapid ASKAP Continuum Survey \citep[RACS;][]{McConnell2020} and the Very Large Array Sky Survey \citep[VLASS;][]{Lacy2020}.
Therefore, in this work, we aim to perform a new search of radio emission in ULXs using these radio archives (\S~\ref{sec:sample}). 
Follow-up observations of some of the selected objects with the Australia Telescope Compact Array (ATCA) were conducted (\S~\ref{sec:atca}). 
The detailed results for each object in the crossmatched sample are shown in \S~\ref{sec:results}. 
We discuss their possible physical nature in \S~\ref{sec:discussion} with a summary in \S~\ref{sec:summary}. 

\section{Sample, observations, and analysis}
\label{sec:sample}

We adopted the ULX catalog of \citet{Walton2022}, which is a compilation of the Chandra Source Catalog Release 2 \citep[CXC2;][]{CXO_catalog2}, the fourth XMM-Newton serendipitous source catalog \citep[4XMM;][]{Webb2020}, and the second Swift X-ray Point Source catalog \citep[2SXPS;][]{Evans2020}, containing 1843 ULXs in nearby galaxies. 
% \st{The position error is provided for each object individually, and the typical 3$\sigma$ radius is about 3\arcsec, 5\arcsec and 10\arcsec\, for Chandra, XMM-Newton and Swift, respectively. } 
The RACS and VLASS radio surveys are used for searching for their radio counterparts.
Two bands of RACS are used, the RACS-low at 0.9 GHz with a bandwidth of 288 MHz and RACS-mid at 1.4 GHz with a bandwidth of 144 MHz. 
The central frequency of VLASS is 3 GHz and the bandwidth is 2 GHz. 
The spatial resolution is in the range of 15\arcsec--25\arcsec\ for RACS-low and $\gtrsim 8$\arcsec for RACS-mid, and the sensitivity is 0.2--0.4~\mjybeam\ and 0.15--0.4~\mjybeam\ for the two bands, respectively \citep{Duchesne2023}. 
For VLASS, the beam size is about 2\farcs5 and the 5$\sigma$ sensitivity is around 0.35~\mjybeam\ \citep{Lacy2020}.
The RACS survey mainly covers the southern sky ($\delta < +40^\circ$) and VLASS mainly covers the northern sky ($\delta > -40^\circ$). 
The combination of the two surveys allows for a full sky coverage at a similar sensitivity of several hundred $\mu$Jy~beam$^{-1}$. 

As the radio counterparts could be rather extended, e.g., a double lobed structure, the search is not based on cross-correlation between the source positions. 
We also realized that some point-like radio sources are not listed in the VLASS source catalog \citep{Gordon2021} and examination of the image is needed.
Thus, we extracted radio images with a dimension of $2\arcmin \times 2\arcmin$ around the ULX position, and try to identify excessive radio emission that is 4 times above the rms noise. 
For an unresolved radio source, the 1$\sigma$ position error is typically the beam size divided by the signal-to-noise ratio, which is used as a rough estimate. 
In sum, we labeled three types of potential counterparts: (i) the radio peak and X-ray position is consistent within the quadratic sum of their 3$\sigma$ position errors; (ii) the X-ray position is associated with an extended radio structure; (iii) the X-ray position is along the center of a double lobed structure no matter if there is a radio core around the X-ray position. 
We selected 21 candidate radio counterparts of ULXs with the above means.
Among them, five are associated with extended radio emission, determined if the integrated flux density deviates from the peak flux density by a factor of 2 or more\footnote{We tested the criterion against a sample of compact radio sources \citep{Petrov2025} with a successful rate of 98\% for RACS and 92\% for VLASS. We also had visual inspections to see whether the sources are resolved or not. We note that these are all empirical and preliminary.}, and three are associated with a double lobed structure with two showing a radio core. 
% \red{To be conservative, we also checked the radio sky survey images of these candidates, which can help for the extended and not extended source classification. }
 
We retrieved optical images around the ULX position from the following archives, prioritized by data quality and availability: the Panoramic Survey Telescope and Rapid Response System \citep[PanSTARRS;][]{panstarrs2016}, Dark Energy Survey Data Release 2 \citep[DES-DR2;][]{Abbott2021}, and Digitized Sky Survey (DSS). 
We note that, limited by the sensitivity and resolution of these ground-based optical surveys, the optical counterparts are usually not due to emission from the ULX binary system \citep{Tao2011} and too bright to be a single star, but could be their host star clusters or host dwarf galaxies, or instead are active galactic nuclei (AGNs) that are masqueraded as ULXs. 
If no optical counterpart can be seen, an upper limit is estimated based on the nearby, faintest objects. 
We note that the optical emission is extended for three sources, NGC 2207 ULX3, NGC 2903 ULX3, and NGC 5457 ULX4, and the optical magnitude represents emission from the bright knot associated with the ULX in the star forming region. 

We argue that the astrometric uncertainty does not affect counterpart identification. 
For the Chandra and XMM catalogs, the systematics have been corrected with residuals included in the quoted uncertainties \citep{CXO_catalog2,Webb2020}. 
For Swift, the statistical errors dominates \citep{Evans2020}. 
The optical images have been astrometrically calibrated to a much higher accuracy than in other bands \citep{Abbott2021, panstarrs_cali}. 
The RACS images may have an astrometric uncertainty of around 2\arcsec, much smaller than the source position uncertainty in our cases \citep{Duchesne2024}. 
For VLASS, the absolute astrometry has an accuracy better than 0\farcs5, negligible again \citep{Gordon2021}. 
Therefore, we conclude that the absolute astrometry is either already corrected or accurate enough compared with statistical uncertainties.

General information of the candidates is listed in Table~\ref{tab:sample}, including the X-ray position and luminosity, distance to the host galaxy, and magnitude of the optical counterpart, etc. 
The radio contours are plotted on top of the optical images in Figure~\ref{fig:img}, overlaid with the ULX position error circle.
The peak flux densities that are directly measured from the survey images are listed in Table~\ref{tab:radio}, along with the spectral index $\alpha$ ($S_\nu \propto \nu^\alpha$) determined using all available bands.

% , X-ray sources have been quoted in Table \ref{tab:sample}, which are conservative enough for the identification of optical counterparts. The systematic errors of RACS-mid are about 1\arcsec--2\arcsec\,~\citep{Duchesne2024} and that of VLASS could be about 0.2\arcsec--0.3\arcsec\,\citep{Gordon2021}, which are about 1 order smaller than their beam size, respectively. There could not be an optical counterpart out the error circle of X-ray source or out the beam size region in Figure \ref{fig:img}. For example, NGC 5457 XMM88 has a VLASS counterpart, which is not spatially consistent with its optical counterpart, which shows that the radio and optical emissions are not from the same region.

%%%%%%%%%%%%%%%%%%%%%%%%%%%%%%%%%%%%%%%%%%%%%%%%%
\begin{deluxetable*}{llcccccclrc}
\tabletypesize\footnotesize
\colnumbers
\tablecaption{ULXs with radio counterparts in the RACS and/or VLASS surveys.}
\label{tab:sample}
\tablewidth{\columnwidth}
\tablehead{ 
\colhead{No.} & \colhead{Name} & \colhead{R.A.} & \colhead{Dec.} & \colhead{$r_{\rm err}$} & \colhead{\emph{D}}  & \colhead{Cat$_{\rm X}$}             & \colhead{$ L_{\rm X}$} &  \colhead{\emph{V}} & \colhead{$\log(f_{\rm X} / f_V)$} & \colhead{Nature} \\ 
\colhead{ } & \colhead{ } & \colhead{(J2000)} & \colhead{(J2000)} & \colhead{(\arcsec)} & \colhead{(Mpc)} & \colhead{ }  & \colhead{($ 10^{40}~{\rm erg~s^{-1}}$)}             & \colhead{ } &  \colhead{  } & \colhead{  } 
}
\startdata 
1 & ESO 352-41 ULX1 & 01 19 07.1 & $-$34 07 05 & 4.0 & 78.4 & 4XMM & $0.1$ & $>23.2$ & $>-0.50$ & \uppercase\expandafter{\romannumeral3}\\  
2 & NGC 925 ULX1 & 02 27 27.5 & $+$33 34 42 & 0.8 & 8.7 & 4XMM & $4.5$ & 20.3 & $1.90$ & \uppercase\expandafter{\romannumeral3}\\  
3 & NGC 1332 ULX2 & 03 26 14.8 & $-$21 21 26 & 1.1 & 24.5 & CSC2 & $0.4$ & 21.6 & $0.75$ & \uppercase\expandafter{\romannumeral1}\\  
4 & PGC 146838 ULX1 & 05 01 07.9 & $-$38 43 21 & 2.8 & 226.7 & 4XMM & $32.0$ & 19.5 & $-0.42$ & \uppercase\expandafter{\romannumeral3}\\  
5 & NGC 2207 ULX3 & 06 16 15.8 & $-$21 22 03 & 1.2 & 36.4 & CSC2 & $3.7$ & 18.2 & $-0.25$ & \uppercase\expandafter{\romannumeral2}\\  
6 & NGC 2903 ULX3 & 09 32 09.8 & $+$21 31 06 & 8.6 & 9.3 & 2SXPS & $0.3$ & 16.6 & $-0.85$ & \uppercase\expandafter{\romannumeral2}\\  
7 & NGC 3190 ULX1 & 10 18 04.3 & $+$21 49 31 & 1.1 & 22.2 & CSC2 & $0.1$ & 24.3 & $1.32$ & \uppercase\expandafter{\romannumeral1}\\  
8 & ESO 171-5 ULX1 & 12 01 32.9 & $-$53 21 36 & 9.3 & 59.9 & 2SXPS & $7.7$ & 16.7 & $-0.99$ & \uppercase\expandafter{\romannumeral3}\\  
9 & NGC 5018 ULX6 & 13 13 06.4 & $-$19 31 14 & 2.2 & 34.0 & 4XMM & $0.2$ & $>21.6$ & $>-0.10$ & \uppercase\expandafter{\romannumeral3}\\  
10 & NGC 5457 ULX4 & 14 02 28.2 & $+$54 16 26 & 1.0 & 7.0 & 4XMM & $0.3$ & 15.7 & $-0.89$ & \uppercase\expandafter{\romannumeral2}\\  
11 & NGC 5457 XMM88 & 14 03 53.9 & $+$54 15 58 & 5.7 & 7.0 & 2SXPS & $0.2$ & 21.5 & $1.18$ & \uppercase\expandafter{\romannumeral3}\\  
12 & ESO 514-3 ULX1 & 15 18 37.1 & $-$24 07 18 & 1.3 & 43.9 & 4XMM & $2.5$ & $>21.0$ & $>0.51$ & \uppercase\expandafter{\romannumeral1}\\  
13 & NGC 5985 ULX1 & 15 39 39.8 & $+$59 19 52 & 1.5 & 52.7 & 4XMM & $8.7$ & 21.1 & $0.93$ & \uppercase\expandafter{\romannumeral3}\\  
14 & NGC 6027 ULX1 & 15 59 10.3 & $+$20 46 19 & 1.1 & 67.6 & CSC2 & $0.2$ & 20.8 & $-0.71$ & \uppercase\expandafter{\romannumeral3}\\  
15 & IC 4587 ULX1 & 15 59 49.5 & $+$25 56 32 & 9.7 & 200.8 & 2SXPS & $63.0$ & 22.7 & $1.30$ & \uppercase\expandafter{\romannumeral3}\\  
16 & UGC 10143 ULX1 & 16 02 18.2 & $+$15 59 12 & 1.5 & 140.6 & CSC2 & $1.5$ & 20.6 & $-0.64$ & \uppercase\expandafter{\romannumeral3}\\  
17 & NGC 6926 ULX1 & 20 33 05.9 & $-$02 02 09 & 3.8 & 88.0 & 4XMM & $2.6$ & $>19.2$ & $>-0.80$ & \uppercase\expandafter{\romannumeral2}\\  
18 & NGC 6946 ULX1 & 20 35 00.7 & $+$60 11 30 & 0.5 & 7.7 & 4XMM & $2.1$ & 19.7 & $1.41$ & \uppercase\expandafter{\romannumeral3}\\  
19 & PGC 90367 ULX1 & 20 37 17.2 & $+$25 31 42 & 1.2 & 153.2 & CSC2 & $1.5$ & $>18.6$ & $>-1.60$ & \uppercase\expandafter{\romannumeral2}\\  
20 & NGC 7059 ULX1 & 21 27 29.2 & $-$60 01 02 & 6.9 & 31.9 & 2SXPS & $31.6$ & 18.1 & $0.77$ & \uppercase\expandafter{\romannumeral3}\\  
21 & NGC 7648 ULX1 & 23 23 53.2 & $+$09 39 59 & 1.3 & 52.2 & CSC2 & $0.5$ & $>17.2$ & $>-1.66$ & \uppercase\expandafter{\romannumeral3}\\
\enddata 
\tablecomments{
Column~(1): object number.
Column~(2): host galaxy and ULX name.
Column~(3): right ascension. 
Column~(4): declination. 
Column~(5): 3$\sigma$ position error of a ULX quoted from \citet{Walton2022}.
Column~(6): distance to the host galaxy adopted from \citet{Walton2022}.
Column~(7): original X-ray catalog from which the position error is quoted.
Column~(8): peak X-ray luminosity in the unit of $10^{40}$~\ergs. 
Column~(9): $V$-band magnitude translated from the photometry in PanSTARRS, DES-DR2, Gaia, or SDSS.
Column~(10): X-ray to optical flux ratio.
Column~(11): possible physical nature of the radio emissions: (\uppercase\expandafter{\romannumeral1}) background quasars, (\uppercase\expandafter{\romannumeral2}) star forming regions, or (\uppercase\expandafter{\romannumeral3})  accreting compact objects in the host galaxy.
}
\end{deluxetable*}
%%%%%%%%%%%%%%%%%%%%%%%%%%%%%%%%%%%%%%%%%%%%%%%%%

\subsection{ATCA observations}
\label{sec:atca}

We conducted radio observations (code: C3663) of eight sources (ESO 352-41 ULX1, NGC 1332 ULX2, PGC 146838 ULX1, NGC 2207 ULX3, ESO 171-5 ULX1, NGC 5018 ULX6, ESO 514-3 ULX1, and NGC 7059 ULX1) in our sample with ATCA at 5.5 GHz and 9 GHz to obtain radio images for a better sensitivity and at higher frequencies.
The array comprises six 22 m antennas with a maximum baseline of 6 km. 
Observations in the 4 cm band were carried out at two central frequencies --- 5.5 GHz and 9 GHz --- each with a 2 GHz bandwidth, using B1934‑638 as the amplitude and band‑pass calibrator.
The data were processed with the Common Astronomy Software Applications \citep{2007ASPC..376..127M}, including flagging of bad data, calibration, Fourier transformation, deconvolution, and image analysis.

\begin{figure}
\centering
\includegraphics[width=0.49\columnwidth]{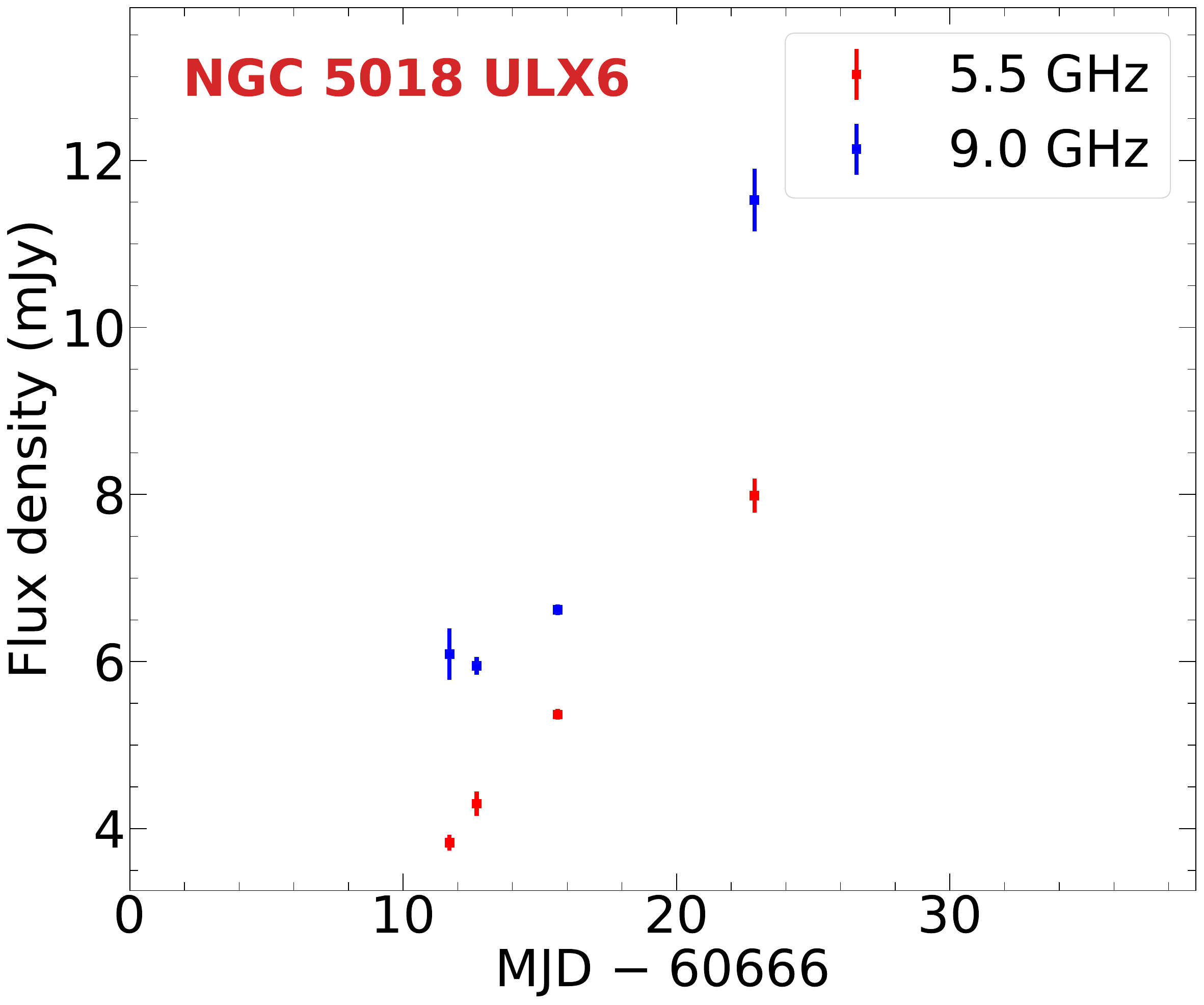}
\includegraphics[width=0.49\columnwidth]{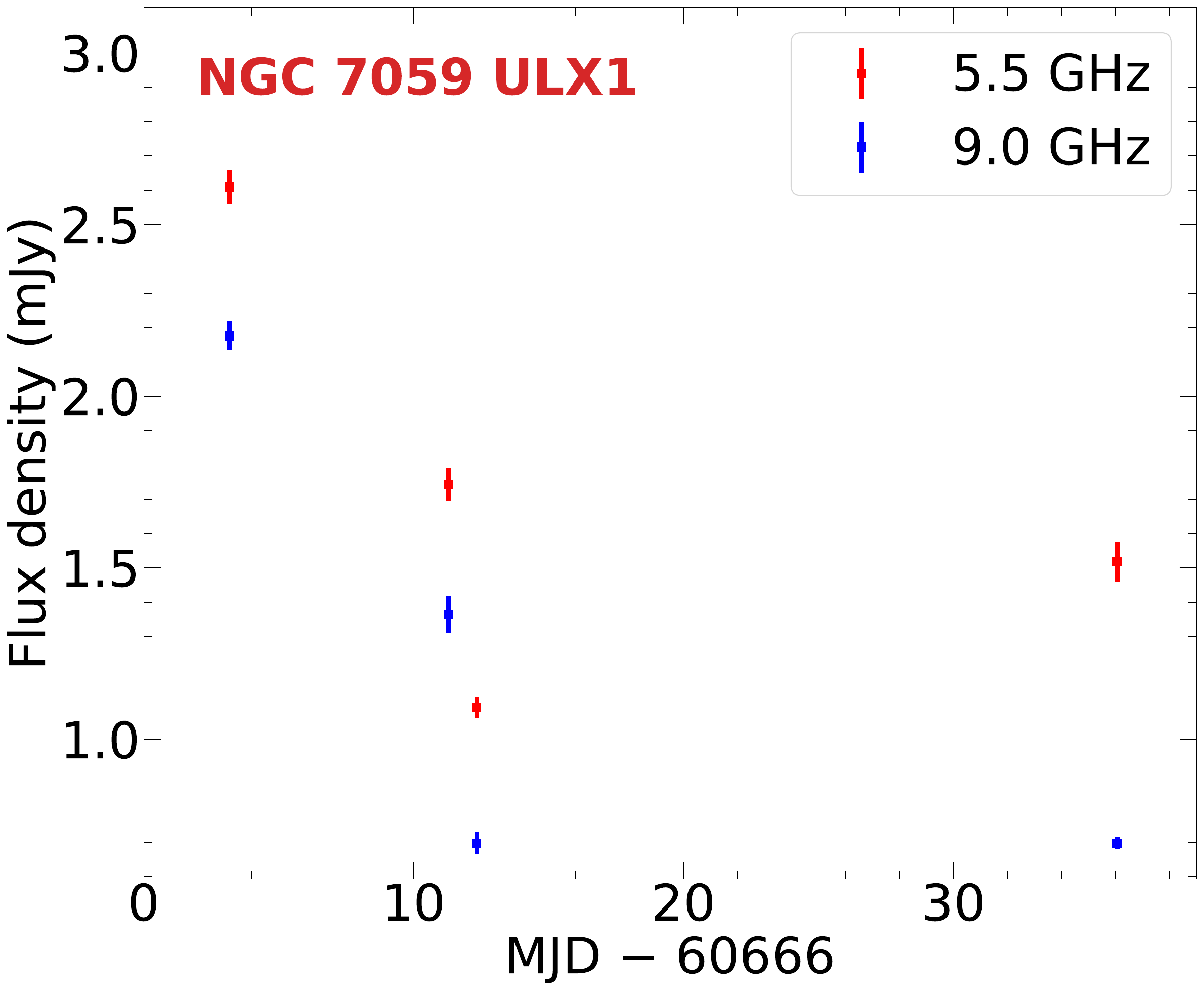}
\caption{ATCA radio lightcurves of NGC 5018 ULX6 and NGC 7059 ULX1 with 1$\sigma$ error bars.}
\label{fig:atca_var}
\end{figure}

The flux contours are also plotted in Figure~\ref{fig:img} along with the survey images.
The peak flux density is listed in Table~\ref{tab:radio}.
The ATCA observation of each target consists of 2--5 individual observations scheduled in different days.
We examined the flux variation and found two sources with time variability.
Their lightcurves are shown in Figure~\ref{fig:atca_var}.

%%%%%%%%%%%%%%%%%%%%%%%%%%%%%%%%%%%%%%%%%%%%%%%%%

\subsection{X-ray to optical flux ratio}

We calculated the X-ray to optical flux ratio following the definition in \citet{Maccacaro1988},
\begin{equation}
\log(f_{\rm X} / f_V) = \log(f_{\rm X}) + \frac{m_V}{2.5} + 5.37 \; ,
\label{eq:flux_ratio}
\end{equation}
where $f_{\rm X}$ is the observed 0.3--3.5 keV flux, and $m_V$ is the $V$-band magnitude. 
The 0.3--3.5 keV flux is transferred from the catalog flux assuming a power-law spectrum subject to interstellar absorption given spectral parameters quoted in the catalog.
The $V$ magnitude is translated from the original survey photometry using the recipes in their references \citep{panstarrs2016, Abbott2021}.
There are a few cases where the survey does not provide the photometry for the object. 
We then adopted the Gaia \citep{GaiaDR3} photometry in one case and Sloan Digitized Sky Survey (SDSS) photometry \citep{Jester2005} in the other two and obtained the $V$ magnitude following the recipes in the above references. 
The flux ratio $\log(f_{\rm X} / f_V)$ are listed in Table~\ref{tab:sample}.

%%%%%%%%%%%%%%%%%%%%%%%%%%%%%%%%%%%%%%%%%%%%%%%%%
\begin{deluxetable*}{lccccccc}
\tabletypesize\footnotesize
\tablecaption{Radio peak flux density and spectral index.}
\label{tab:radio}
\tablewidth{\columnwidth}
\tablehead{
\colhead{Name} & \multicolumn{2}{c}{RACS} & \colhead{VLASS} & \multicolumn{2}{c}{ATCA} & \colhead{$\alpha$} & \colhead{Remark} \\
\cline{2-3} \noalign{\smallskip} \cline{5-6} \noalign{\smallskip}
\colhead{} &  \colhead{0.9 GHz} &  \colhead{1.4 GHz}  &  \colhead{3 GHz}             &  \colhead{5.5 GHz}  &  \colhead{9 GHz} & \colhead{} & \colhead{} \\ 
\colhead{ } & \colhead{(mJy)} & \colhead{(mJy)} & \colhead{(mJy)} & \colhead{(mJy)} & \colhead{(mJy)} & \colhead{} & \colhead{}
}
\startdata 
ESO 352-41 ULX1 & $1.4\pm0.2$ & $0.78\pm0.15$ &  & $0.121\pm0.012$ & $0.103\pm0.013$ & $-1.24\pm0.08$ & \\  
NGC 925 ULX1 & $2.2\pm0.4$ & $1.6\pm0.3$ & $0.60\pm0.13$ &  &  & $-1.1\pm0.2$ & \\  
NGC 1332 ULX2 &  &  &  &  &  &  & lobes\\  
PGC 146838 ULX1 &  &  & $0.96\pm0.13$ & $0.86\pm0.04$ & $0.864\pm0.016$ & $-0.03\pm0.08$ & \\  
NGC 2207 ULX3 & $25.0\pm0.8$ & $7.7\pm0.7$ & $2.66\pm0.13$ & $2.09\pm0.04$ & $1.029\pm0.016$ & $-1.260\pm0.016$ & extended\\  
NGC 2903 ULX3 & $12.2\pm0.4$ & $5.9\pm0.9$ & $1.03\pm0.16$ &  &  & $-2.00\pm0.12$ & extended\\  
NGC 3190 ULX1 &  & $1.2\pm0.2$ & $1.48\pm0.14$ &  &  & 0.3 & core+lobes\\  
ESO 171-5 ULX1 & $1.1\pm0.2$ & $0.82\pm0.15$ &  & $0.23\pm0.05$ & $0.150\pm0.009$ & $-0.88\pm0.07$ & \\  
NGC 5018 ULX6 & $1.2\pm0.3$ & $2.18\pm0.16$ & $2.1\pm0.2$ & $4.38\pm0.03$ & $6.21\pm0.06$ & $0.70\pm0.02$ & \\  
NGC 5457 ULX4 &  &  & $0.75\pm0.11$ &  &  &  & extended\\  
NGC 5457 XMM88 &  &  & $1.47\pm0.11$ &  &  &  & \\  
ESO 514-3 ULX1 &  &  &  & $0.34\pm0.04$ & $0.40\pm0.03$ & 0.3 & core+lobes\\  
NGC 5985 ULX1 &  &  & $0.60\pm0.13$ &  &  &  & \\  
NGC 6027 ULX1 & $10.3\pm0.5$ & $7.3\pm0.2$ & $5.07\pm0.12$ &  &  & $-0.52\pm0.04$ & \\  
IC 4587 ULX1 & $152.4\pm0.3$ & $107.2\pm0.2$ & $64.35\pm0.14$ &  &  & $-0.702\pm0.002$ & \\  
UGC 10143 ULX1 & $8.5\pm0.5$ & $6.1\pm0.2$ & $2.9\pm0.2$ &  &  & $-0.88\pm0.07$ & \\  
NGC 6926 ULX1 &  &  & $1.14\pm0.17$ &  &  &  & extended\\  
NGC 6946 ULX1 &  &  & $0.76\pm0.18$ &  &  & $-$0.5\dag & \\  
PGC 90367 ULX1 &  &  & $1.19\pm0.14$ &  &  &  & extended\\  
NGC 7059 ULX1 & $2.6\pm0.2$ & $1.8\pm0.2$ &  & $1.335\pm0.012$ & $0.698\pm0.018$ & $-0.61\pm0.02$ & \\  
NGC 7648 ULX1 &  &  & $0.57\pm0.14$ &  &  &  & \\
\enddata 
\tablecomments{The error of flux is the noise rms and the error of $\alpha$ is at the 1$\sigma$ level. Extended sources indicate that the integrated flux density is higher than the peak flux density by a factor of 2 or more.}
\tablenotetext{^\dag}{Quoted from \citet{Lacey1997}.}
\end{deluxetable*}
%%%%%%%%%%%%%%%%%%%%%%%%%%%%%%%%%%%%%%%%%%%%%%%%%

%%%%%%%%%%%%%%%%%%%%%%%%%%%%%%%%%%%%%%%%%%%%%%%%%
\begin{figure*}
\includegraphics[width=0.245\linewidth]{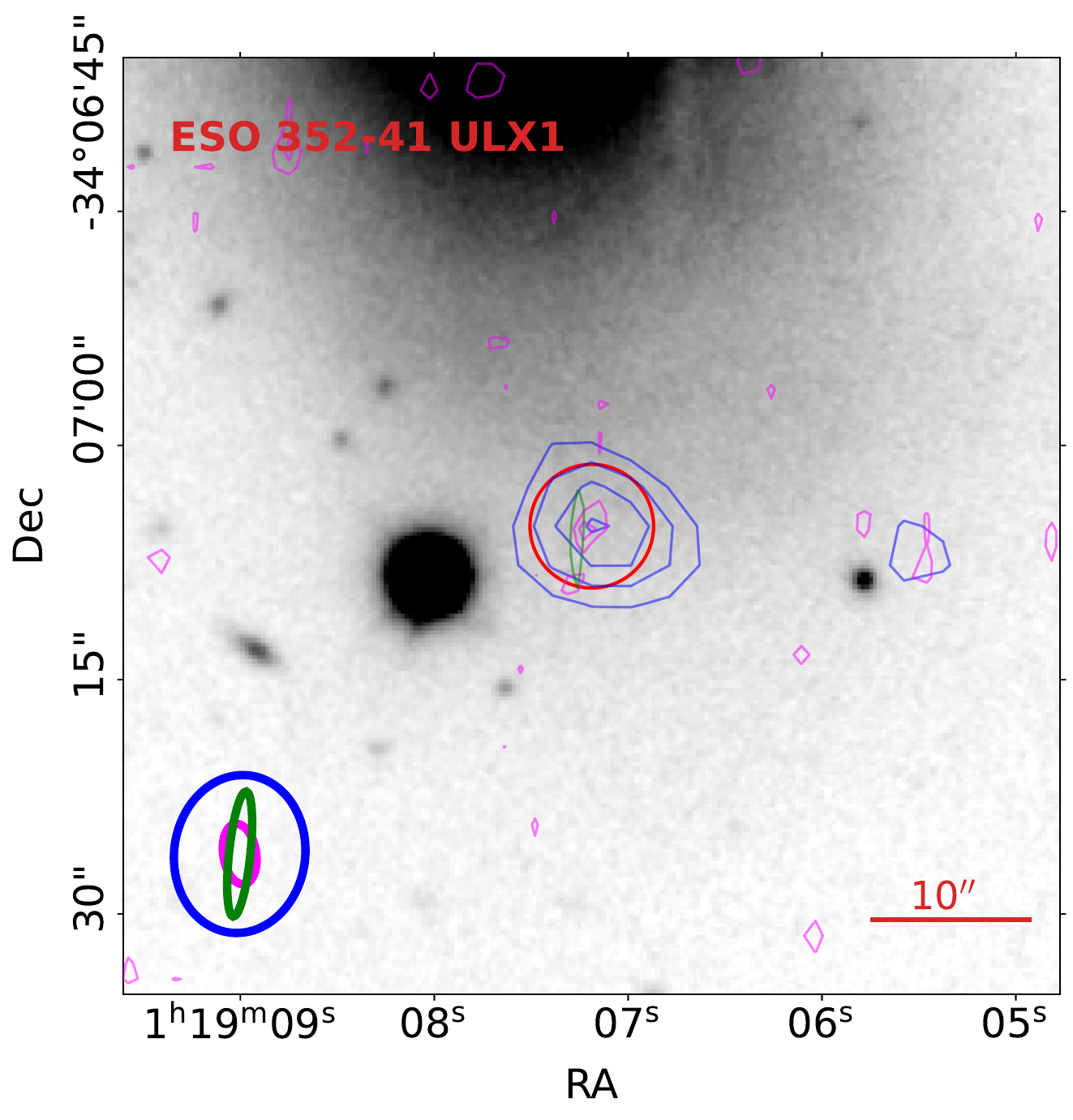}
\includegraphics[width=0.245\linewidth]{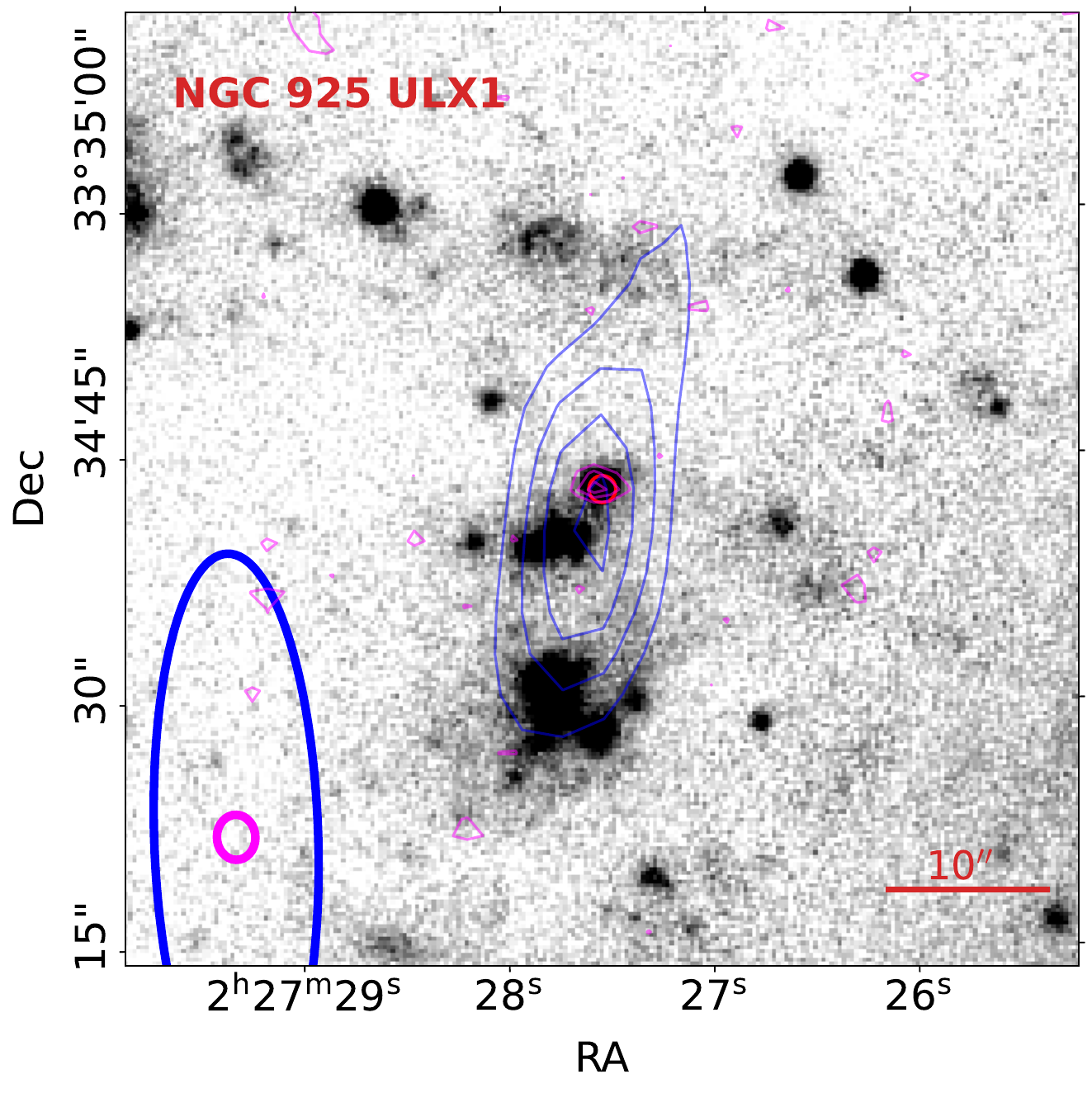}
\includegraphics[width=0.245\linewidth]{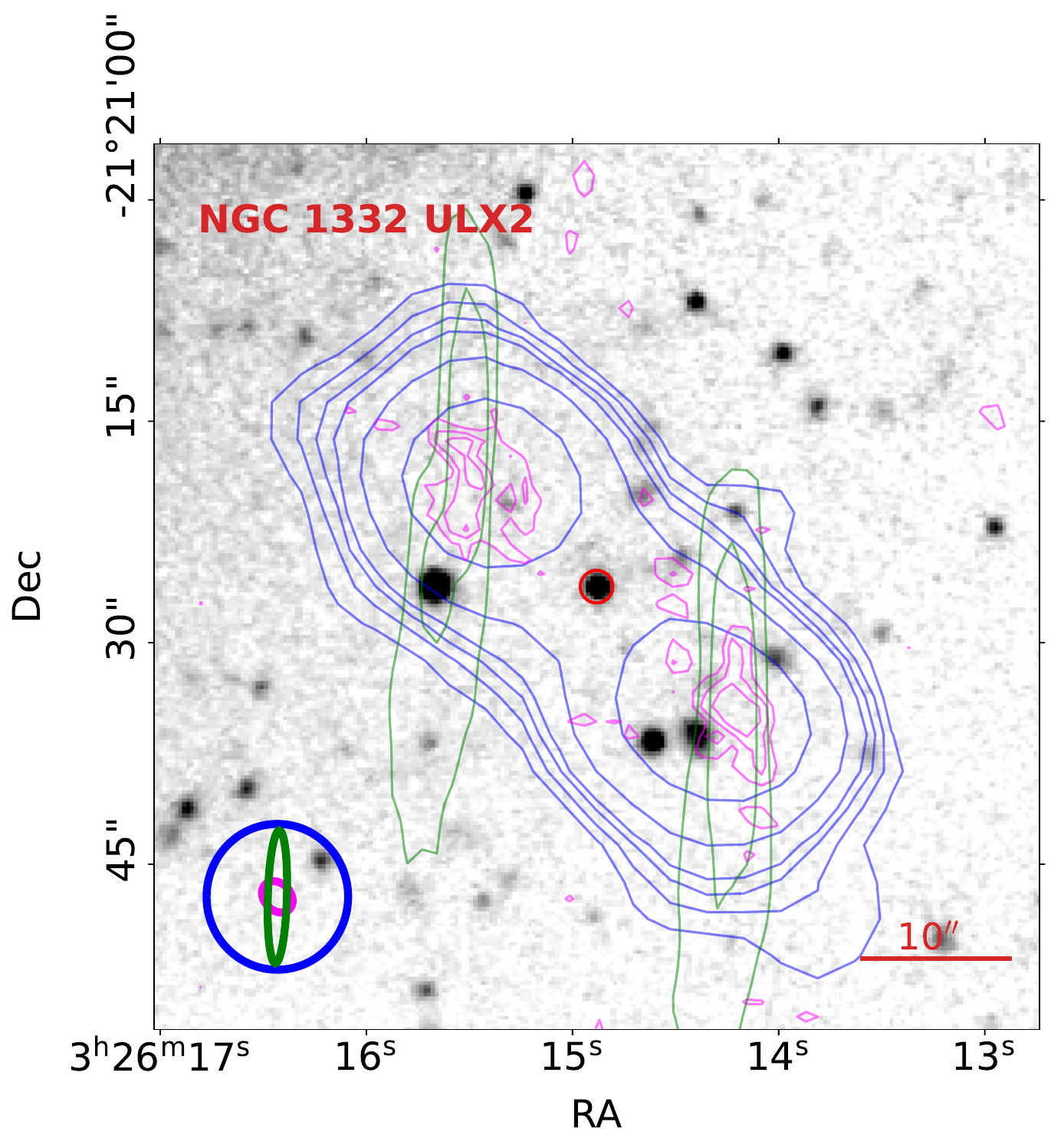}
\includegraphics[width=0.245\linewidth]{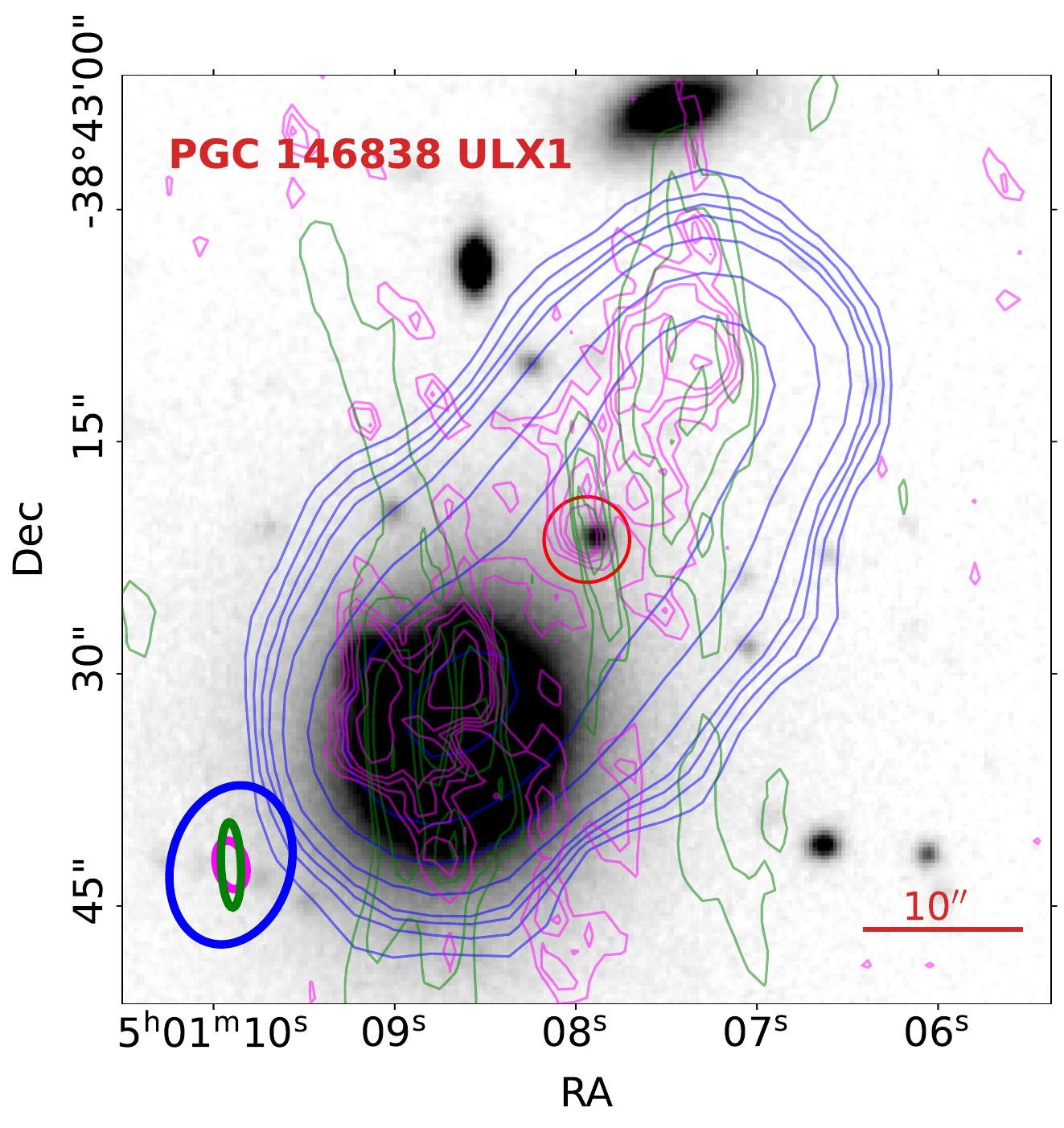}
\includegraphics[width=0.245\linewidth]{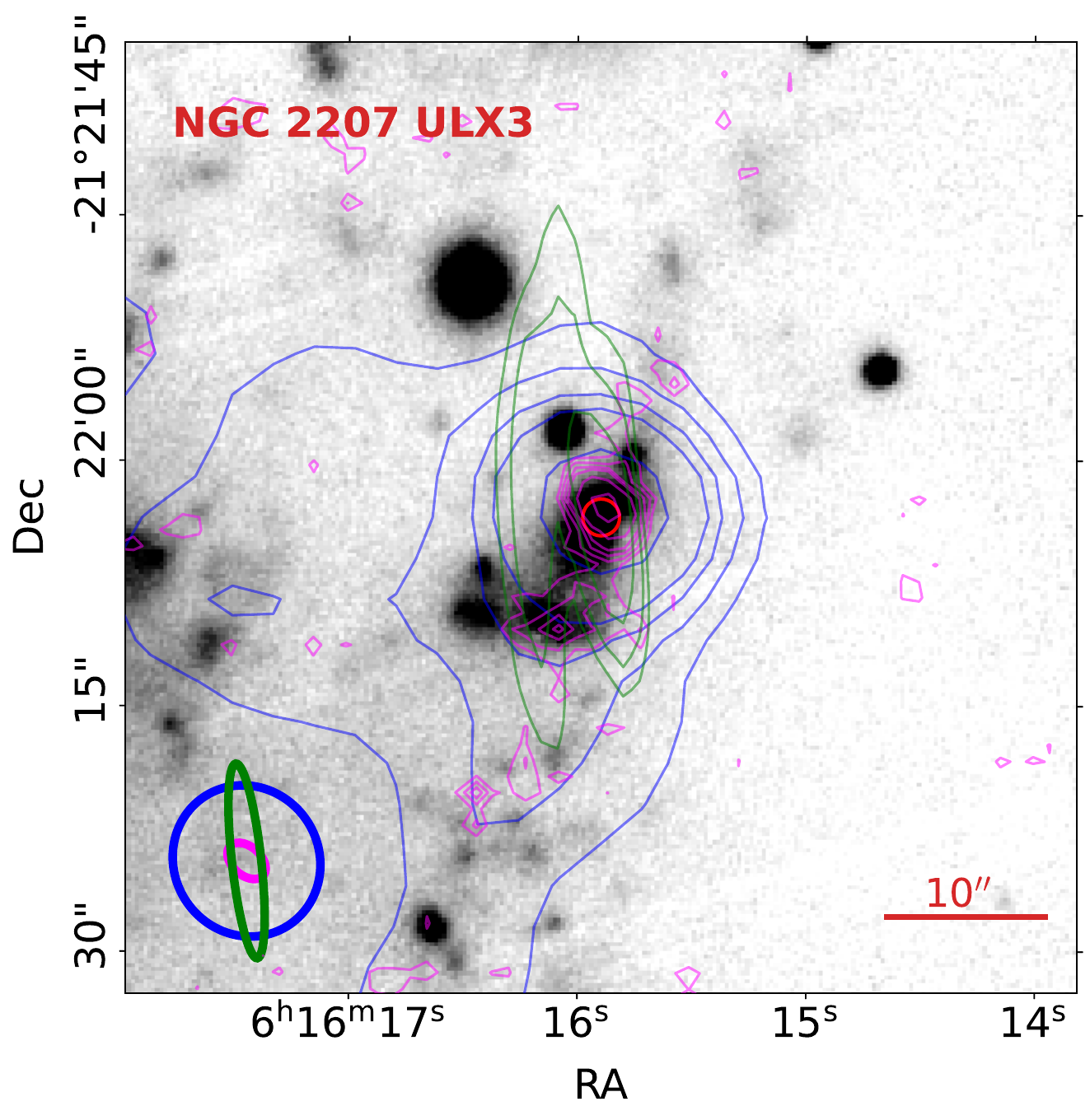}
\includegraphics[width=0.245\linewidth]{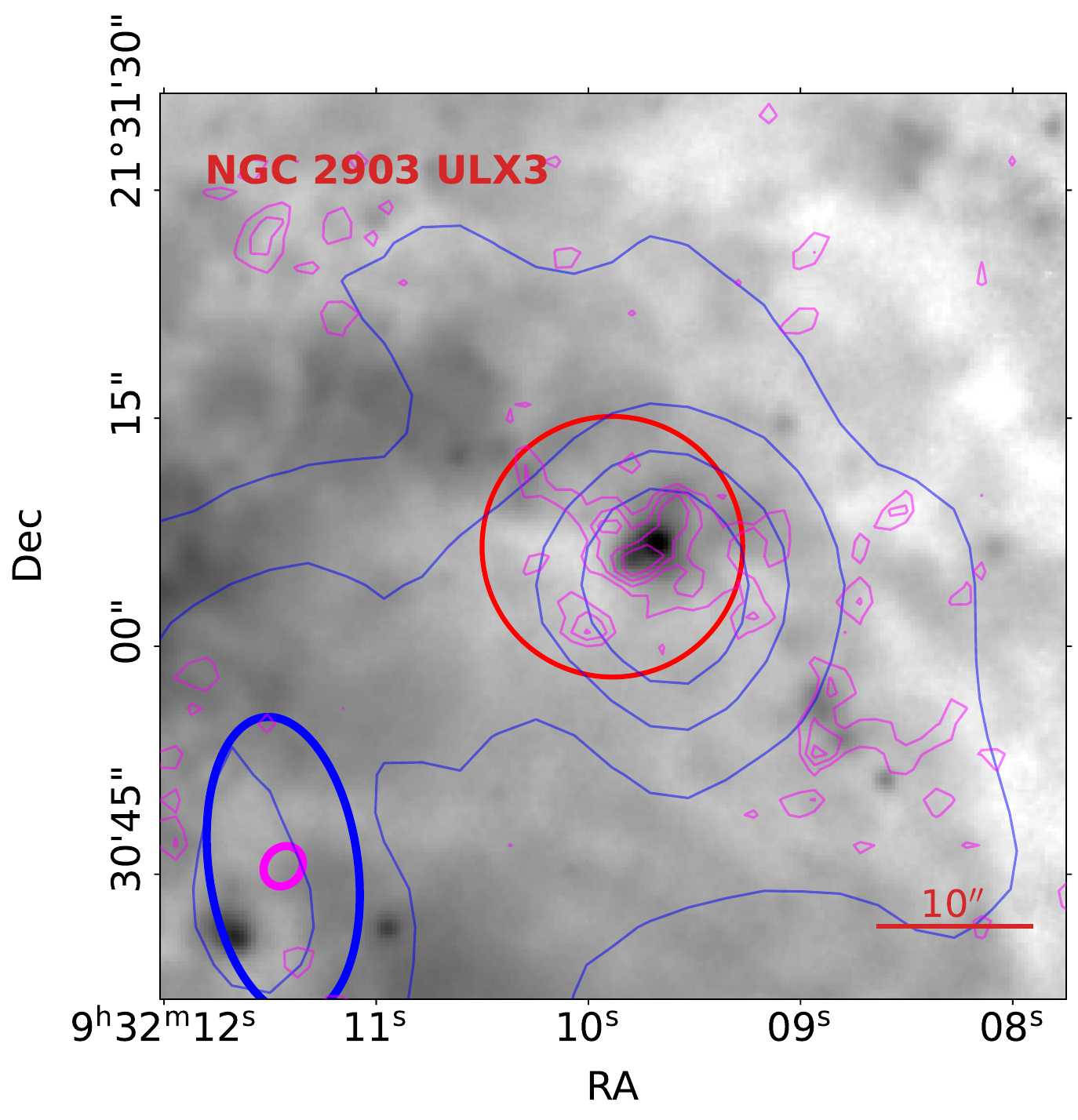}
\includegraphics[width=0.245\linewidth]{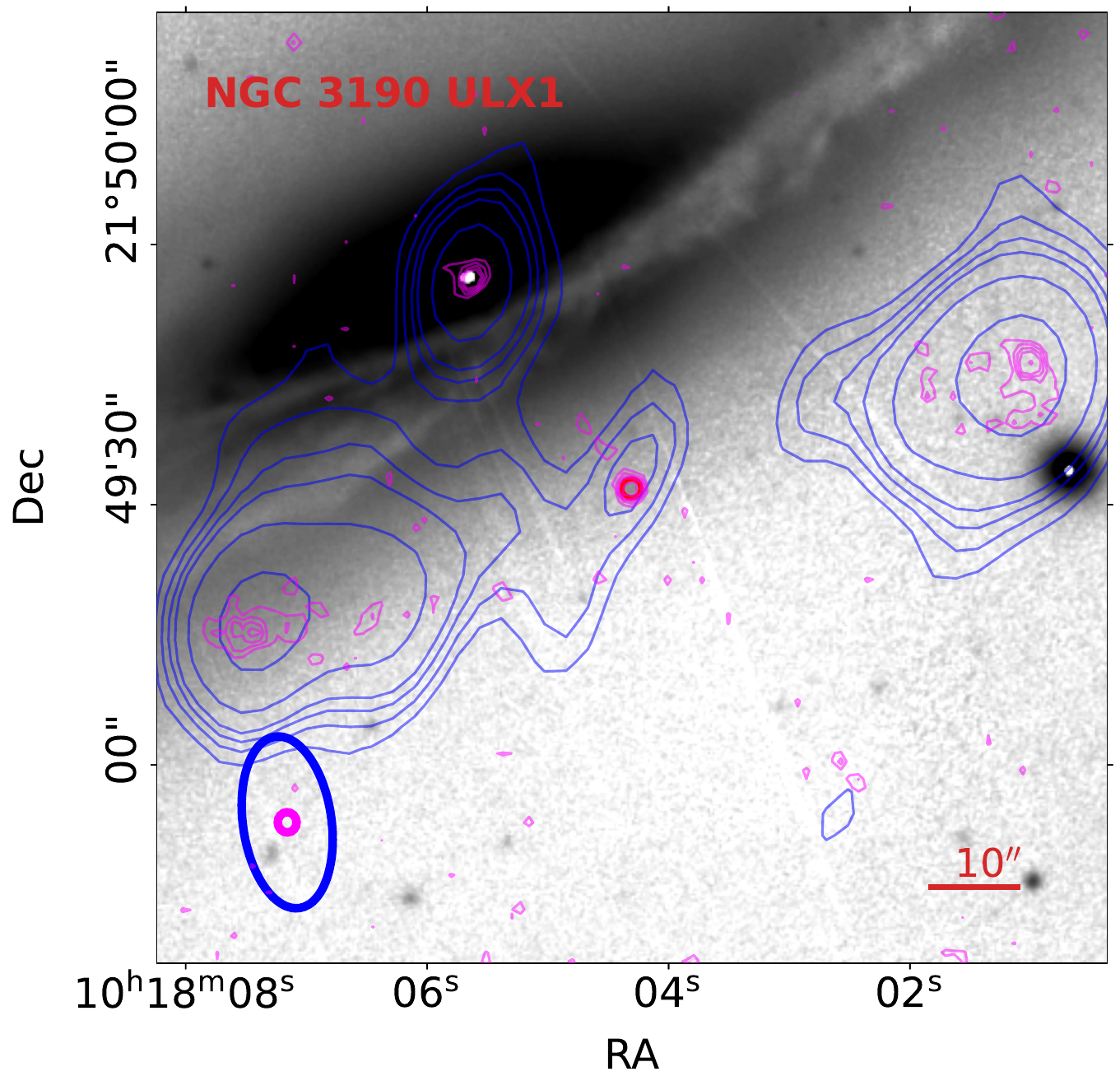}
\includegraphics[width=0.245\linewidth]{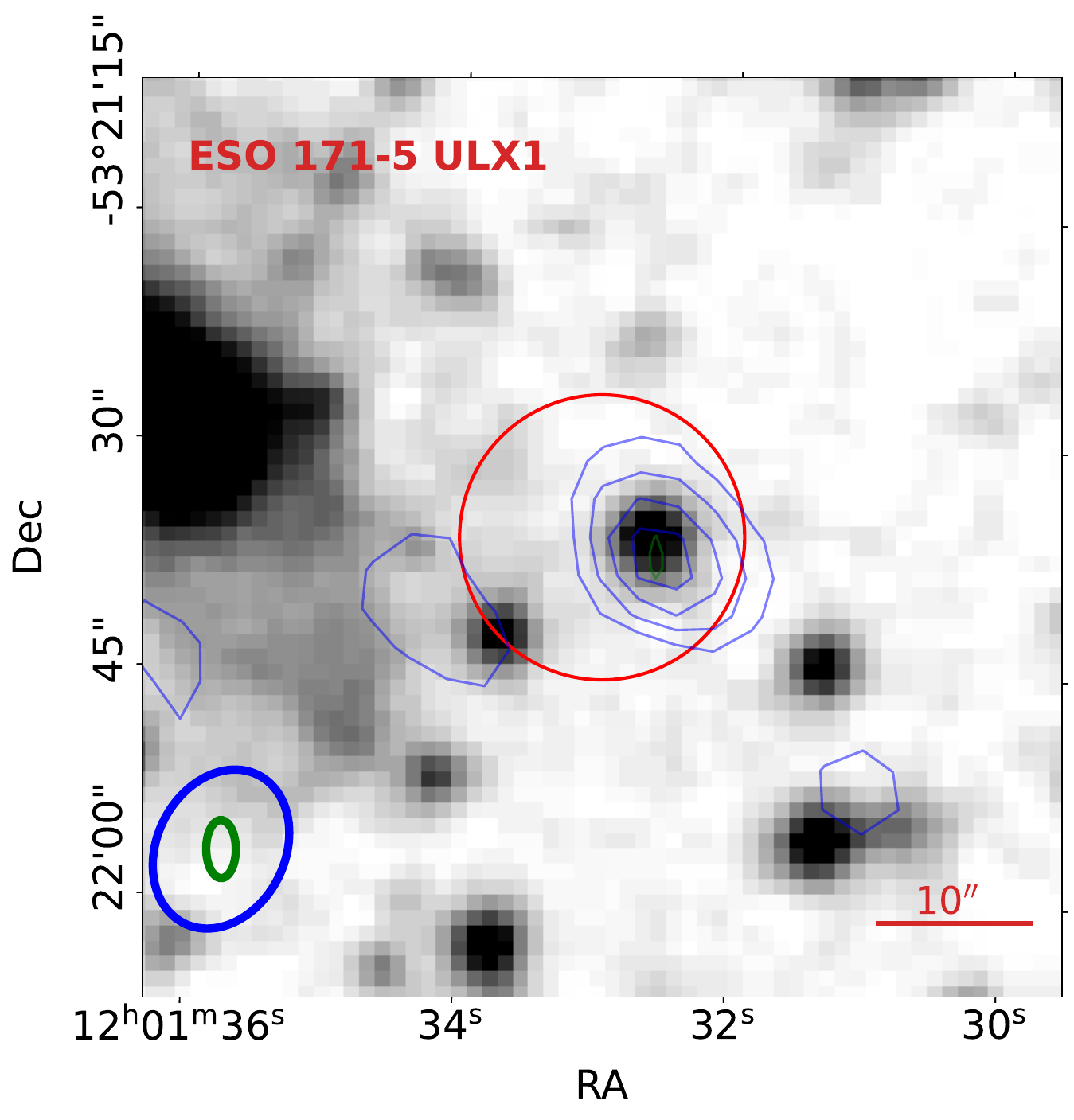}
\includegraphics[width=0.245\linewidth]{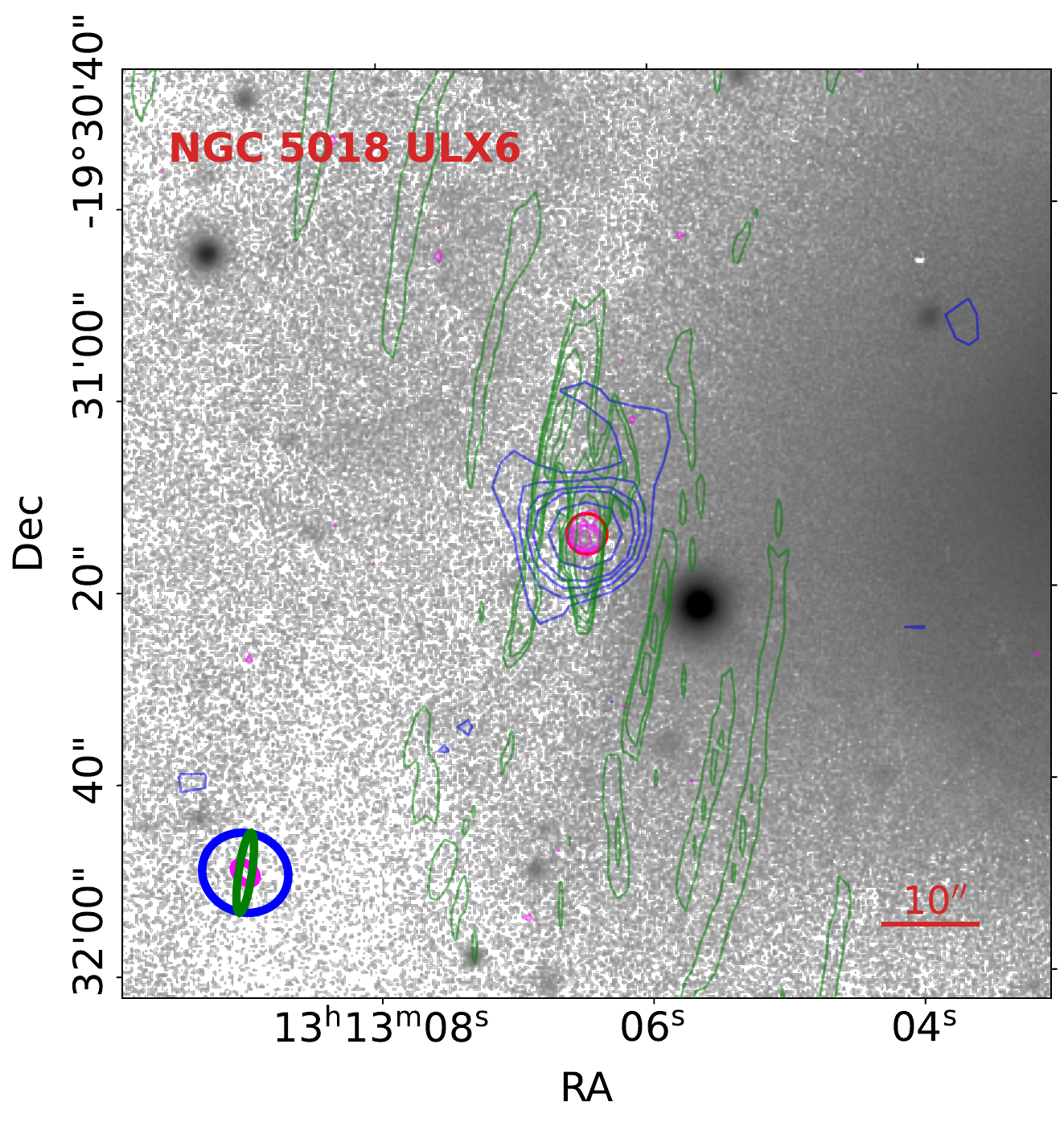}
\includegraphics[width=0.245\linewidth]{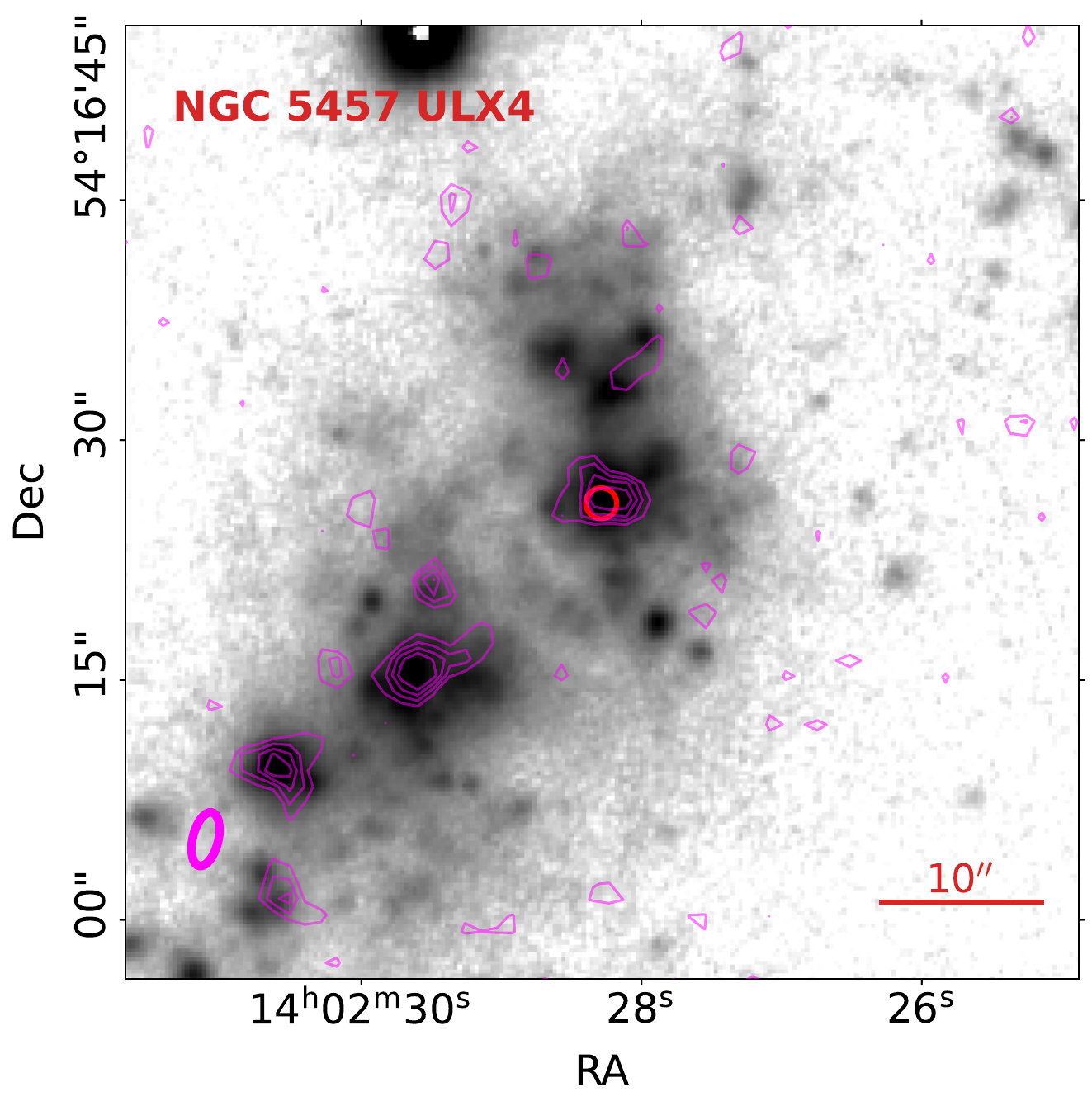}
\includegraphics[width=0.245\linewidth]{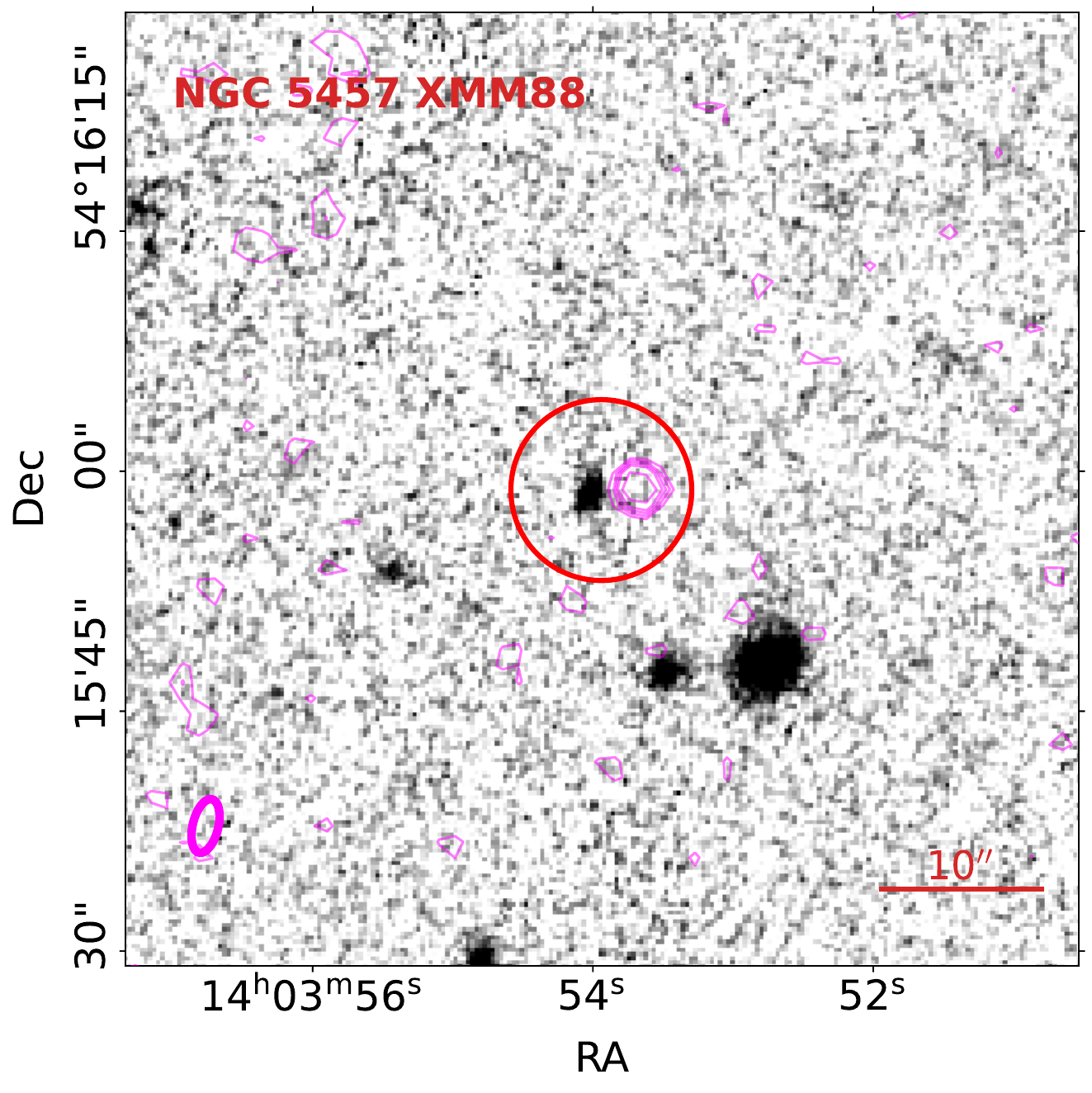}
\includegraphics[width=0.245\linewidth]{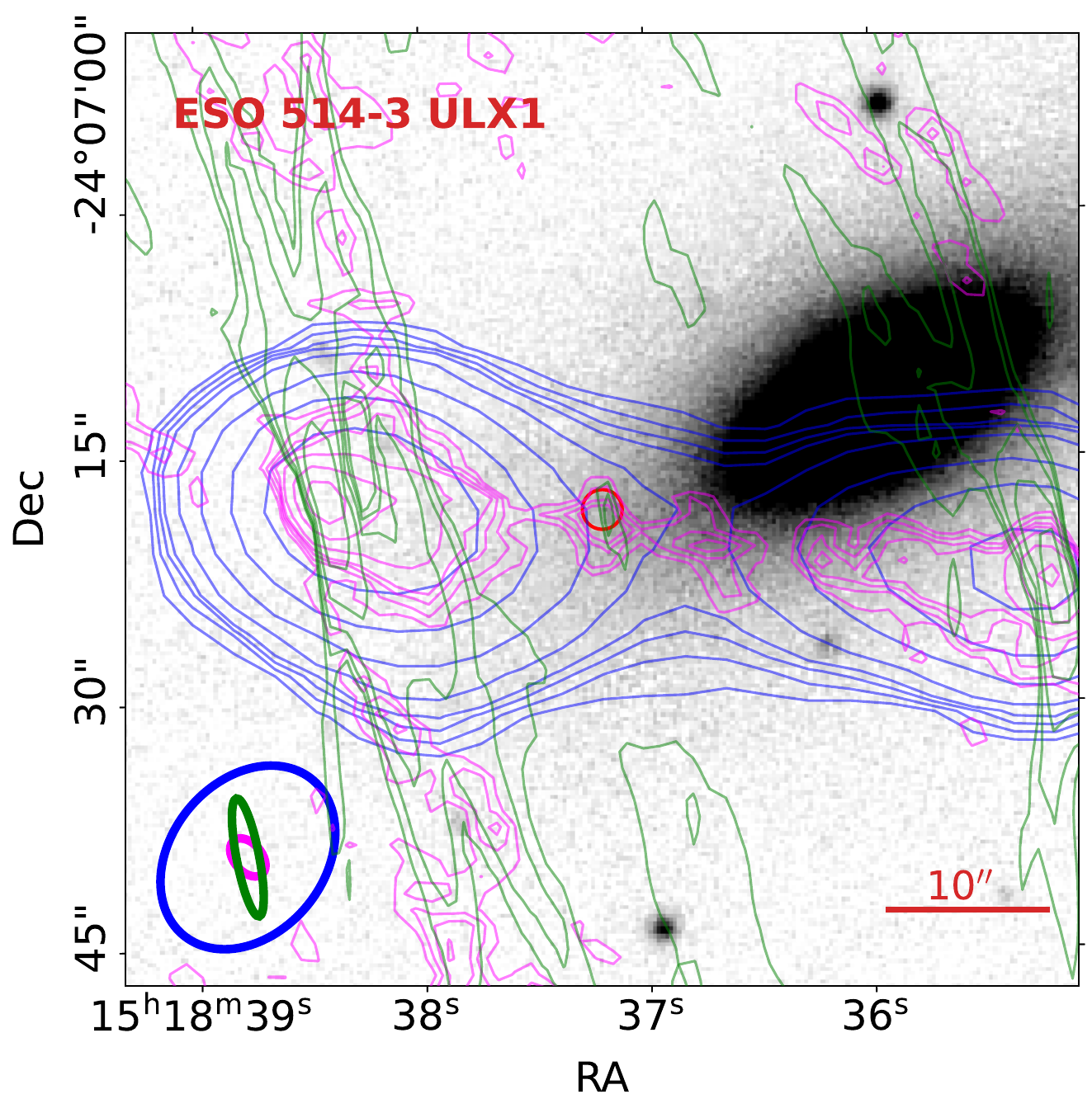}
\caption{Radio contours around the ULX position on top of optical images. 
The red circle indicates the ULX position with a 3$\sigma$ error radius. 
The contours are 2, 3, 4, 5, 8, 16, 32, and 64 times the rms noise level for RACS (blue) and VLASS (magenta), and 8, 16, 32, 64 and 128 times the rms for ATCA (green). 
The beam sizes are shown at the bottom left corner.
The scale bar has a length of 10\arcsec.}
\label{fig:img}
\end{figure*}
%%%%%%%%%%%%%%%%%%%%%%%%%%%%%%%%%%%%%%%%%%%%%%%%%

%%%%%%%%%%%%%%%%%%%%%%%%%%%%%%%%%%%%%%%%%%%%%%%%%
\begin{figure*}
\centering
\setcounter{figure}{0}
\includegraphics[width=0.245\linewidth]{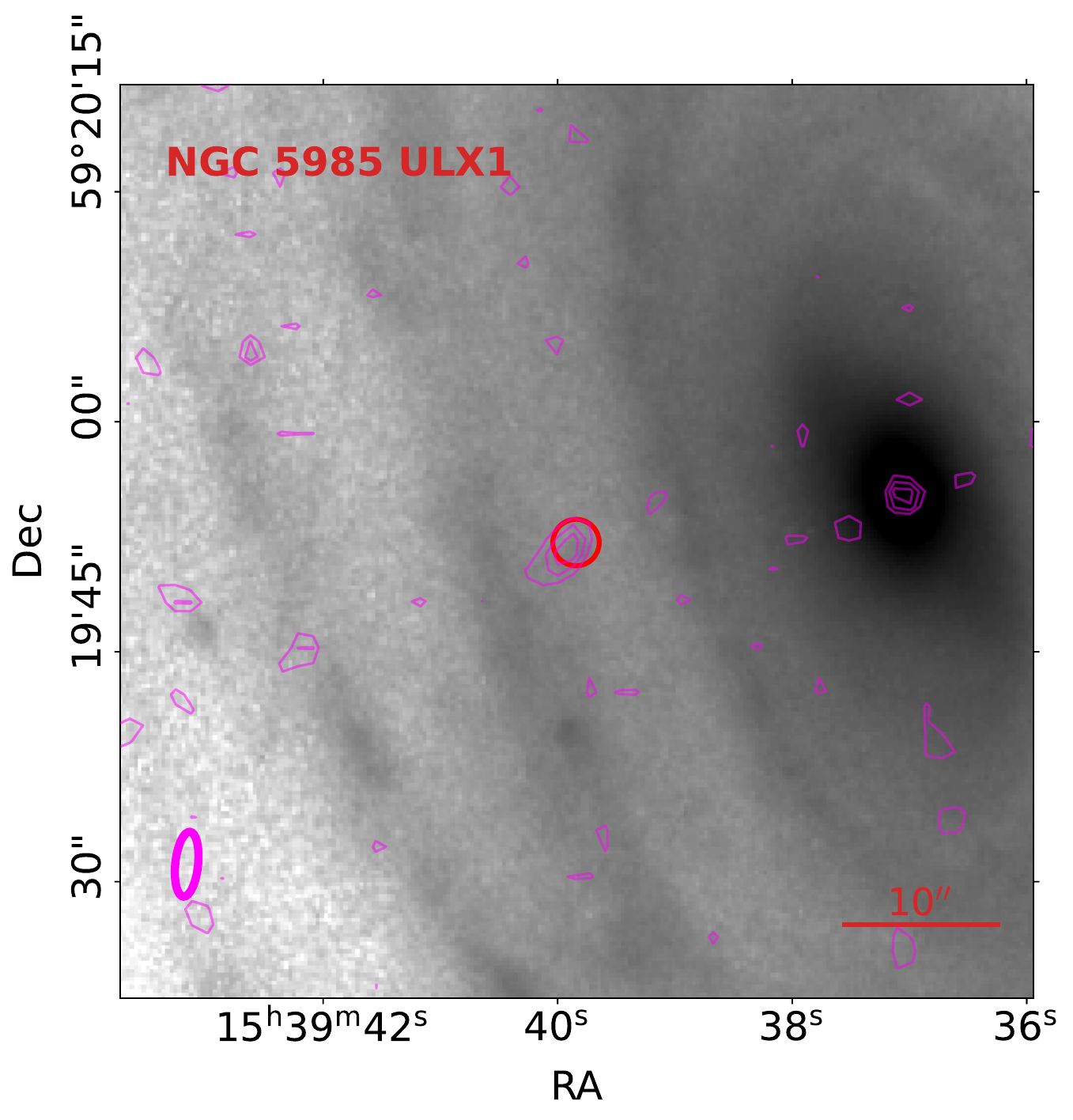}
\includegraphics[width=0.245\linewidth]{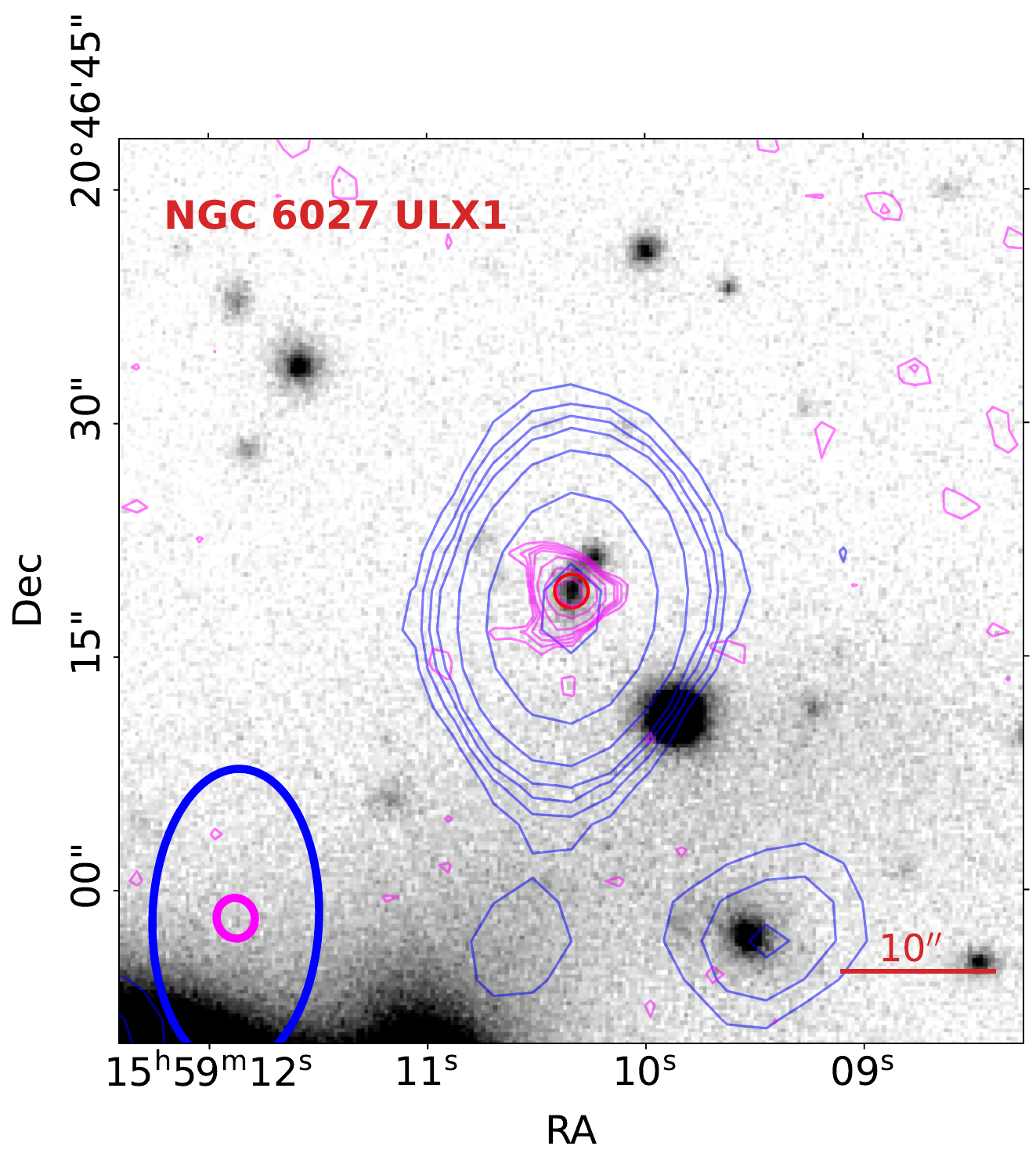}
\includegraphics[width=0.245\linewidth]{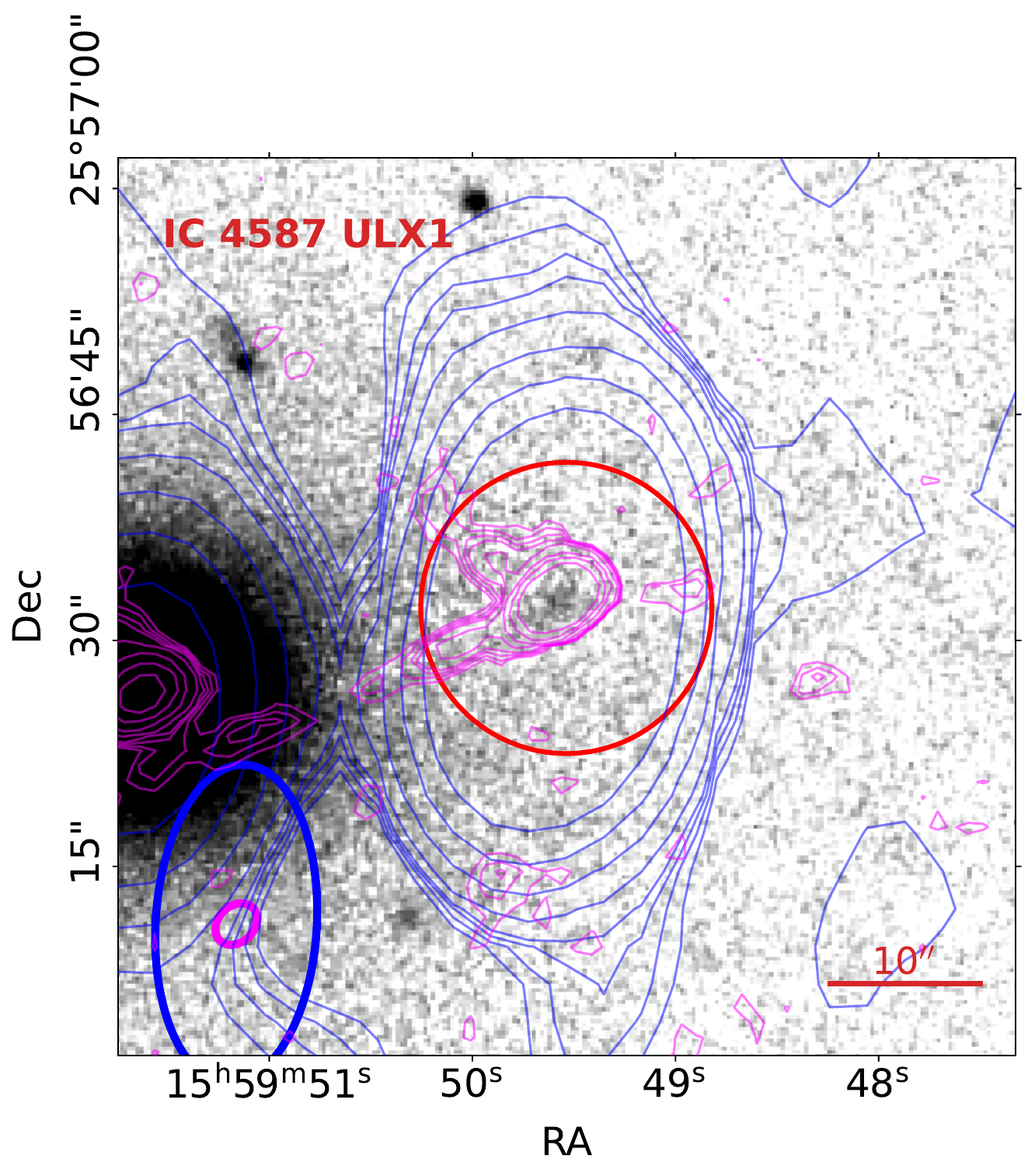}\\
\includegraphics[width=0.245\linewidth]{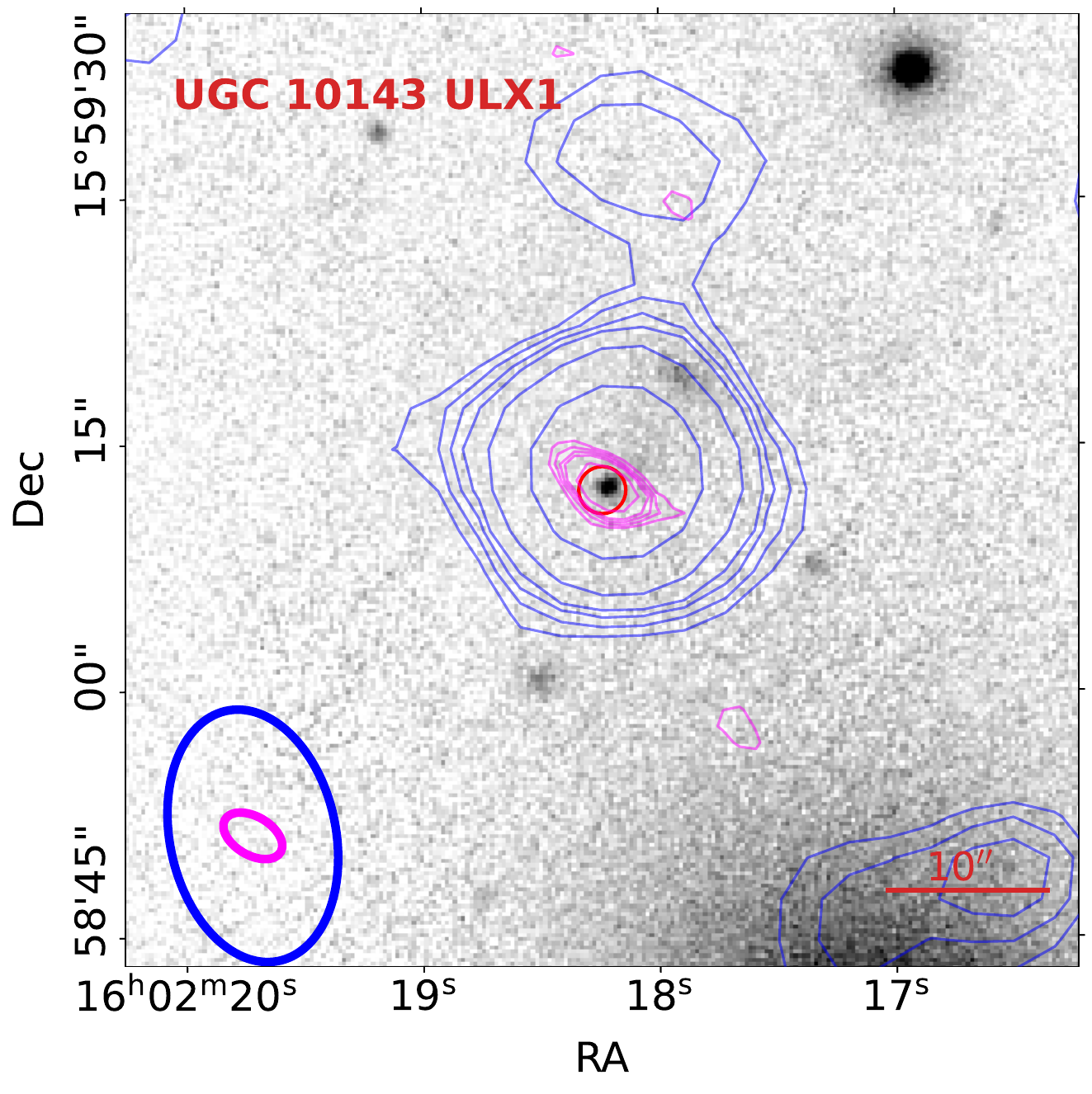}
\includegraphics[width=0.245\linewidth]{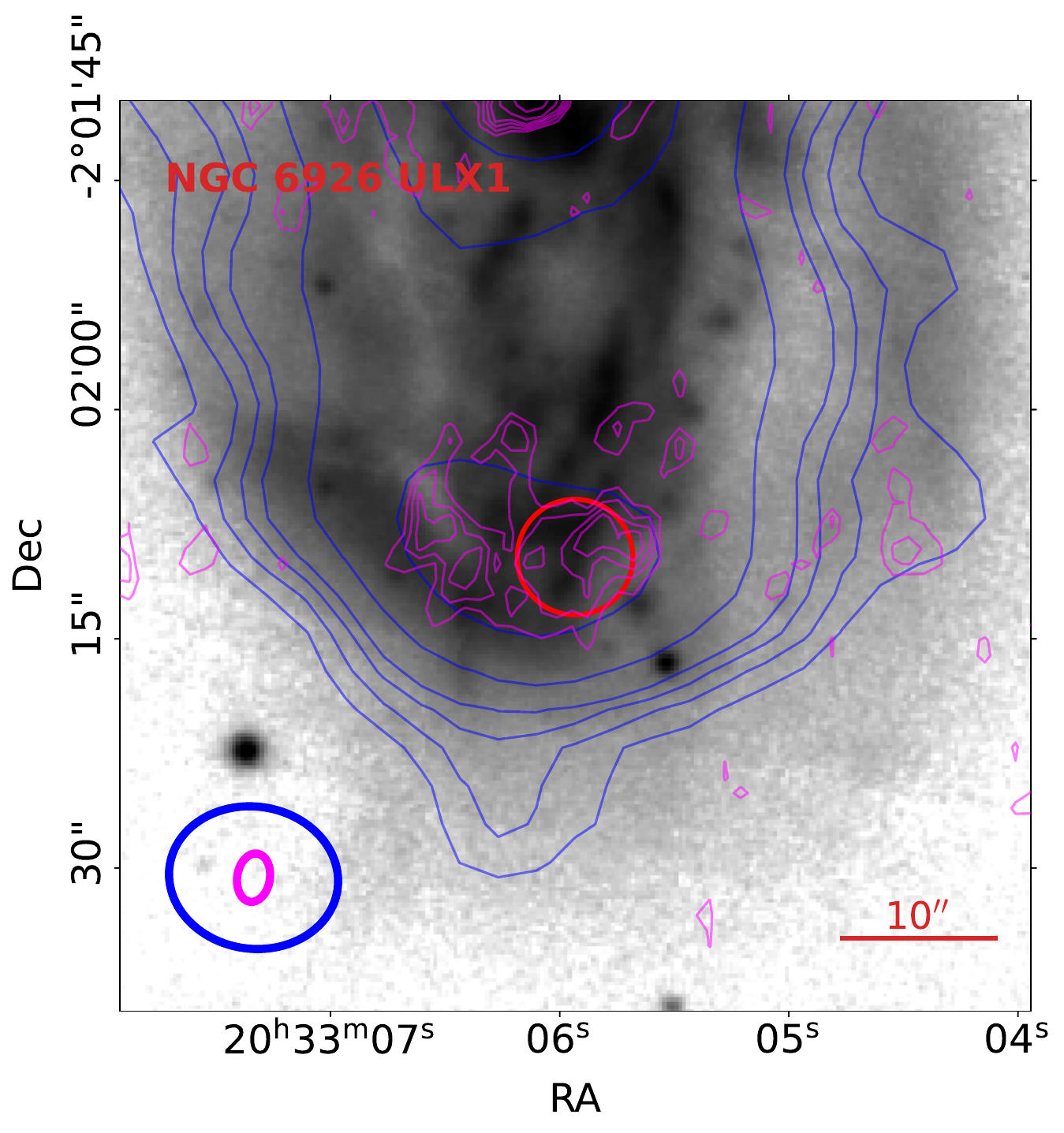}
\includegraphics[width=0.245\linewidth]{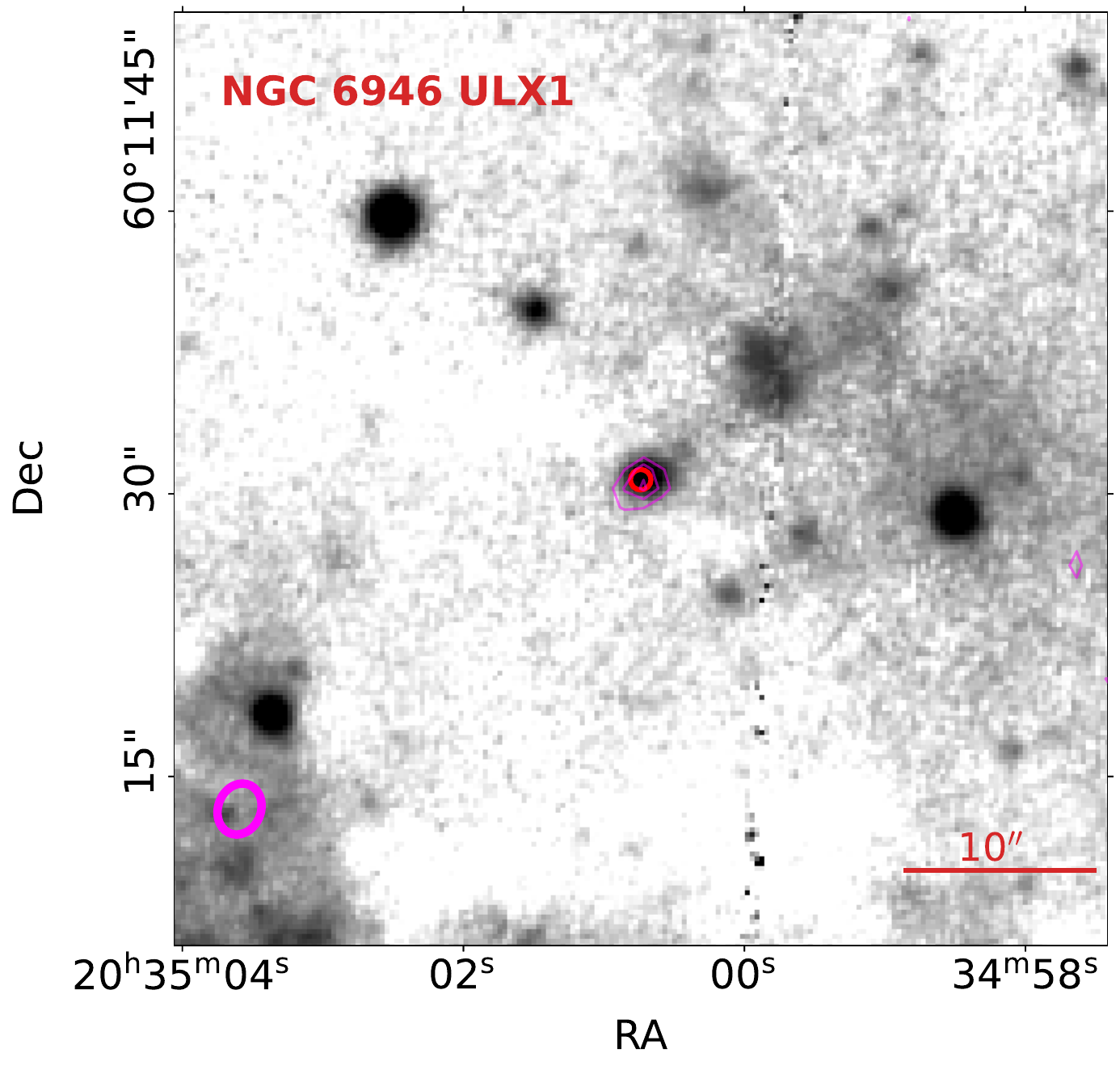}\\
\includegraphics[width=0.245\linewidth]{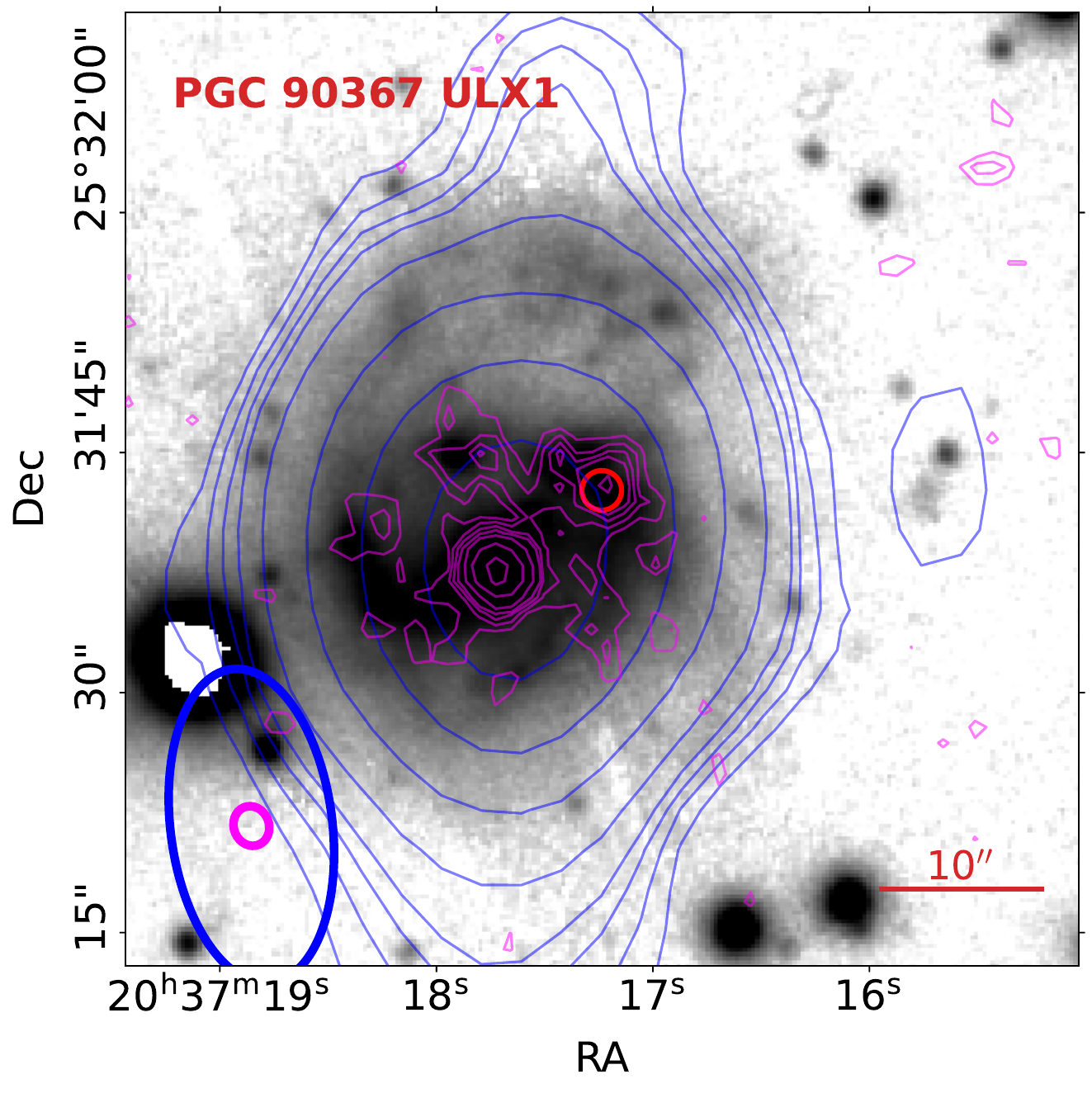}
\includegraphics[width=0.245\linewidth]{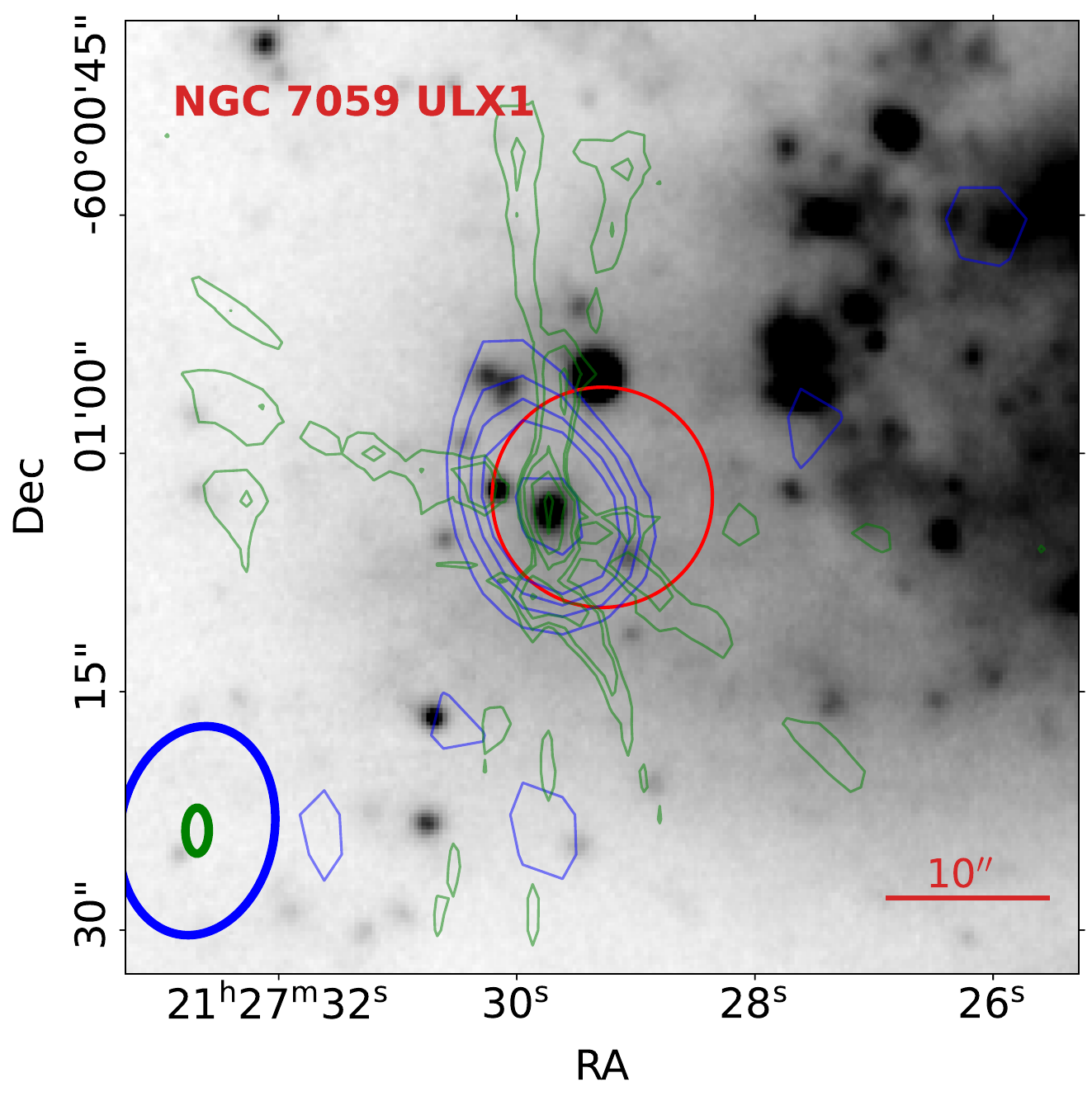}
\includegraphics[width=0.245\linewidth]{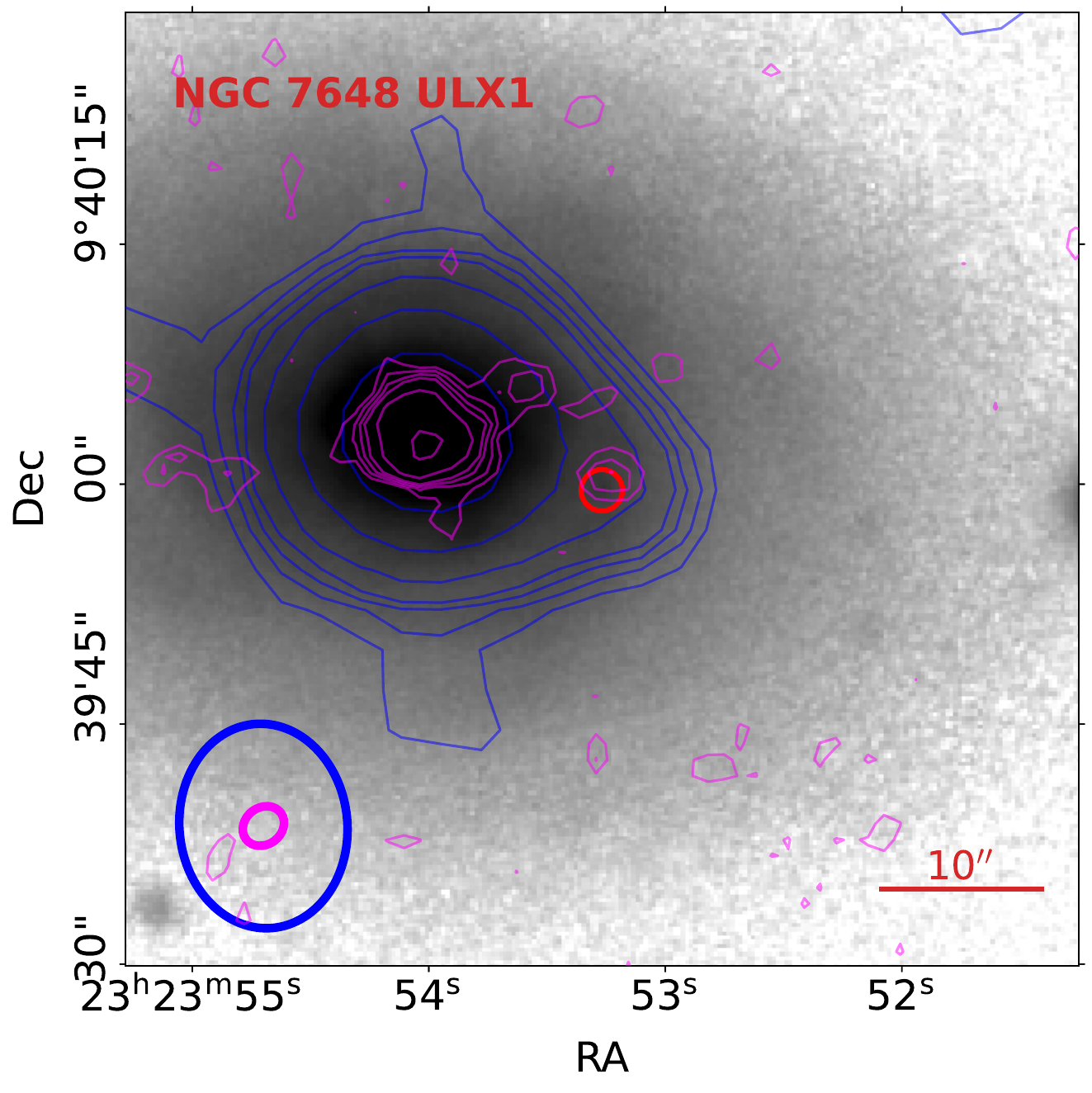}
\caption{Continued.}
\end{figure*}
%%%%%%%%%%%%%%%%%%%%%%%%%%%%%%%%%%%%%%%%%%%%%%%%%

%%%%%%%%%%%%%%%%%%%%%%%%%%%%%%%%%%%%%%%%%%%%%%%%%
%%%%%%%%%%%%%%%%%%%%%%%%%%%%%%%%%%%%%%%%%%%%%%%%%

\section{Results}
\label{sec:results}

In the following we describe details of each object in the sample including related information in the literature.

\subsection{ESO 352-41 ULX1}

There is a spatially unresolved radio source at the ULX position, detected in RACS. 
However, no radio source is detected in VLASS above 4$\sigma$ with an rms of 0.14~mJy at 3~GHz. 
The source was detected with ATCA on 2025 Dec 22 and 25, respectively, at both 5.5 GHz and 9 GHz.
No optical counterpart can be seen in the DES image, with an upper limit of 23.1~mag on the $r$-band. 

\subsection{NGC 925 ULX1}
\label{sec:ngc925}

An unresolved radio source in VLASS is detected at the ULX position.
The ULX displays a high luminosity varying in the range of $(1-5) \times 10^{40}$~\ergs\ with weak spectral variability \citep{Pintore2018, Salvaggio2022}. 
The spectrum of its infrared counterpart shows no continuum but is dominated by an \FeII\ emission line plus many other weaker emission lines, likely due to shocks harder than typical supernova remnants or X-ray illumination of surrounding materials \citep{Heida2016}.
\citet{Lara2021} revealed a large (285~pc diameter) optical emission line nebula around the source with low metallicity, in agreement with the results reported in \citet{Heida2016}.

\subsection{NGC 1332 ULX2}

This is the second brightest X-ray source in NGC 1332 and is outside the D25 isophotal diameter of the galaxy \citep{Humphrey2004}. 
The radio images show a double lobed structure but no compact radio core is detected at the ULX position. 
ATCA observations confirmed the results.   
A point-like optical object at the ULX position can be seen in the DES image. 

\subsection{PGC 146838 ULX1}
 
The RACS image shows an elongated structure with the ULX near the center.  
This structure is obviously attached to the host galaxy, and resolved into two parts in the VLASS image, one around the host galaxy and the other rooted at the ULX position.
Interestingly, ATCA observations identified a point-like radio core at the ULX position with a flat spectrum ($\alpha \approx 0$); the VLASS image at the ULX position also shows an enhanced flux. 
A point-like optical counterpart is seen in the DES image. 

\subsection{NGC 2207 ULX3}

This is source 28 in \citet{Mineo2013}, source 3 in \citet{Mineo2014} and NGC 2207-1 in \citet{Smith2014}, spatially coincident with a starburst region called \textit{feature i} \citep{Elmegreen2000}. 
The authors also pointed out that the X-ray source is extended and has a soft spectrum, suggesting that both the X-ray and radio emission could be due to starburst activity. 
However, \citet{Mineo2014} reported distinct X-ray fluxes measured from the source with four Chandra observations, varying in the range of $(0.8 - 6) \times 10^{-14}$~\ergcms, suggesting that there could be a ULX embedded in an extended component. 

\subsection{NGC 2903 ULX3}

It is the third most luminous X-ray source in NGC 2903 \citep{Perez-Ramirez2010}. 
The ULX is spatially coincident with an extended \HII\ region \citep[source 36 with a radius of 6\farcs4 in][]{Popping2010}. 
The radio emission is extended seen in the VLASS image.
The VLASS radio contour seems to match the optical morphology in the PanSTARRS image. 

\subsection{NGC 3190 ULX1}

A bright radio core plus two lobes are clearly detected with both RACS and VLASS. 
A faint optical counterpart is seen in the PanSTARRS image. 

\subsection{ESO 171-5 ULX1}

The ULX has an unresolved radio counterpart in RACS.
Within the X-ray position uncertainty (3$\sigma$ radius 9\farcs3), there is a GAIA source with a parallax distance of $2.2_{-0.2}^{+0.4}$~kpc~\citep{Bailer2021}.
The color and magnitude suggest that it is G-type main sequence star. 
Given the GAIA source density around the ULX and the X-ray position error, a chance coincidence with such a bright GAIA source in the field is about 0.1 and thus cannot be ruled out. 
Therefore, a more precise X-ray position is needed to confirm or reject the association.  

\subsection{NGC 5018 ULX6}

The ULX shows an extremely hard X-ray spectrum
with an optical counterpart identified in HST images with a magnitude of about 24 in F555W and F814W \citep[see source 6 in][]{Ghosh2005}.
With the HST magnitudes, we derived an X-ray to optical flux ratio of 0.9.
\citet{Ghosh2005} reported the detection of a radio source at the ULX position, with a flux of 2.1~mJy at 1.4~GHz and an inverted spectrum ($\alpha \approx 0.2$), consist with the survey results in this work. 
They argued that it is a compact core of a background radio galaxy.
Here our ATCA observations detected significant variation in the radio band. 

\subsection{NGC 5457 ULX4}

The ULX is spatially coincident with NGC 5447 \citep{Gordon2008}, an \HII\ region in M101 (= NGC 5457). 
The ULX is cataloged as source 4 in \citet{Jenkins2004}, source 20 in \citet{Jenkins2005}, X17 in \citet{Liu2005}, XMM4 in \citet{Winter2006}, and C21 in \citet{Berghea2008}, and could be an X-ray binary embedded in the \HII\ region. 
The VLASS image shows a small degree of extension compared with the beam pattern. 
\citet{Gajovic2025} reported the radio emission from this \HII\ region. 
The spectrum is a power-law with a slope $\alpha = -0.34$, without spectral flattening (absorption) down to 54~MHz, suggesting that the radio emission may arise from unresolved, separate small \HII\ regions. 
We note that the ULX has displayed both short and long-term variability \citep{Jenkins2005} and significant emission above 2 keV \citep{Winter2006}, indicating that it is an X-ray binary instead of emission from the hot gas.

\subsection{NGC 5457 XMM88}

The ULX is located at the outskirt of M101 (= NGC 5457) on the sky plane. 
The source is labeled as source 88 in \citet{Jenkins2005} from an XMM-Newton observation, during which the X-ray luminosity is only $1.3 \times 10^{38}$~\ergs.
In the Chandra source catalog \citep{CXO_catalog}, the source flux is $2.85 \times 10^{-14}$~\ergcms, corresponding to a luminosity of $1.3 \times 10^{38}$~\ergs\ at a distance of 7~Mpc.
However, the source was caught as a ULX with a Swift observation on 2011 December 6 at a luminosity of $2 \times 10^{39}$~\ergs. 
The Swift detection has a $3\sigma$ position uncertainty of 5\farcs7, consistent with the Chandra and XMM-Newton positions. 
If the X-ray emission detected from the three telescopes are due to the same source, then the Chandra position indicates that the X-ray position does not match the optical source in the PanSTARRS image. 
The radio source is a point-like object in the VLASS image, spatially coincident with the Chandra position. 

\subsection{ESO 514-3 ULX1}

ESO 514-3 is a member of the NGC 5903 group.
It has been reported by \citet{OSullivan2018} that the X-ray source is associated with a compact radio core plus a pair of radio lobes, based on which the authors argued that it is a background AGN. 
They reported a core radio flux of about 1~mJy at 4.8 GHz, which is almost 2-3 times higher than our ATCA measurement.
We note that the total radio flux densities including the core and the two lobes are consistent between the two works, suggesting that there may be temporal variability in the core radio emission.
The optical counterpart is not visible in the PanSTARRS image, with an upper limit estimated to be $V = 21.0$, suggesting that \XOR\ $> 0.5$.

\subsection{NGC 5985 ULX1}

The ULX appears at a sky position between two spiral arms, spatially coincident with an unresolved radio source in the VLASS image. 

\subsection{NGC 6027 ULX1}

NGC 6027 is a member of the Seyfert's Sextet (HCG 79), and the X-ray source is spatially coincident with one of a pair, apparently interacting galaxies \citep{Tamburri2012}.
With HST imaging, \citet{Palma2002} suspected that this object was a background spiral galaxy but no redshift measurement is available. 

\subsection{IC 4587 ULX1}

The X-ray source is near a reddish optical object around the elliptical galaxy IC 4587. 
The radio source is included in the VLBA calibrator survey \citep{Petrov2021}, who obtained a jet orientation (source elongation) of $145\arcdeg \pm 5\arcdeg$ by directly analyzing the calibrated
visibilities. 

\subsection{UGC 10143 ULX1}

The X-ray source appears near the elliptical radio galaxy UGC 10143 and has an optical counterpart. 
The radio source in both VLASS and RACS is unresolved. 

\subsection{NGC 6926 ULX1}

The X-ray source is around the southern spiral arm of the Seyfert 2 Galaxy NGC 6926. 
The radio emission is extended as seen in VLASS, matching a star forming region in optical. 

\subsection{NGC 6946 ULX1}

This is a well studied ULX in NGC 6946, embedded in an optical and radio nebula MF16 \citep{VanDyk1994, Matonick1997, Lacey1997, Beuchert2024}. 
The radio flux density measured with VLASS is consistent with that reported previously with VLA observations \citep[source 85 in][]{Lacey1997}, with which the spectral index was measured as $\alpha = -0.5 \pm 0.1$.  
The radio emission is spatially resolved with VLA observations \citep{VanDyk1994, Beuchert2024}.

\subsection{PGC 90367 ULX1}

The X-ray source is around the spiral arm of PGC 90367. 
The radio emission in the VLASS image appears to be extended, with enhancement at the ULX position. 

\subsection{NGC 7059 ULX1}

NGC 7059 is a spiral galaxy and unique for its association with a Fermi gamma-ray source 4FGL J2127.6$-$5959 \citep{Abdollahi2020}. 
The radio source has a point-like optical counterpart in the DES image.
Therefore, it is likely that the radio, X-ray and gamma-ray emission is due to a background blazar. 
The ATCA observations detect significant variation in the radio band.

\subsection{NGC 7648 ULX1}

The ULX appears on the S0 galaxy NGC 7648 with ongoing star formation \citep{Rose2001}. 
Due to the brightness of the galaxy emission, no obvious optical counterpart can be identified in the PanSTARRS image. 
We note that the radio source associated with the ULX is not on the radial spokes of the nearby AGN. 

\section{Discussion}
\label{sec:discussion}

Here we discuss their possible physical nature in three categories depending on the radio morphology and multiwavelength properties. This suggests that this sample has a diverse population.

\subsection{Double lobed structures and background quasars}

There are three objects where a double lobed radio structure is revealed around the ULX. 
They are NGC 1332 ULX2, NGC 3190 ULX1, and ESO 514-3 ULX1. 
Among them, NGC 3190 ULX1 and ESO 514-3 ULX1 also show a radio core associated with the ULX position.
Multiband radio data suggest that they have a flat spectrum, reminiscent of a compact radio jet. 
A naive speculation is that these are not ULXs, but background quasars masqueraded as ULXs as they appear along the line of sight of a nearby galaxy. 
Based on ground-based optical survey images, NGC 1332 ULX2 and NGC 3190 ULX1 show a point-like optical counterpart, allowing us to estimate the X-ray-to-optical flux ratio \XOR\ $=$ 0.75 and 1.32, respectively. 
The ratios are consistent with those of AGNs or BL Lac objects \citep{Maccacaro1988}. 
For ESO 514-3, its X-ray-to-optical flux ratio ($> 0.5$), as well as the possible core radio flux variation, is consistent with the properties of an AGN or blazar. 
Following \citet{Landt2006}, we calculated $L_{\rm core} / L_{\rm 1keV}$ and $L_{\rm core} / L_{\rm lobe}$, where $L_{\rm core}$ and $L_{\rm lobe}$ are the core and lobe luminosity densities at 1.4 GHz, respectively, and $L_{\rm 1keV}$ is the X-ray luminosity density at 1 keV, and found that the three sources all populate in the lobe-dominated regime and follow the correlation found for radio quasars. 

Assuming that they follow the fundamental plane of accreting black holes, we may estimate the black hole mass using the relation in \citet{Merloni2003}. 
The flux density at 5 GHz is estimated from the best-fit radio power-law spectrum. 
The inferred black hole mass is $5 \times 10^{7}$~$M_\sun$ for NGC 3190 ULX1 and $4 \times 10^{6}$~$M_\sun$ for ESO 514-3 ULX1, again consistent with the scenario that they are background quasars. 
To firmly determine their nature, a redshift measurement is needed.

Assuming that the three objects are background quasars, it sets a lower limit of the contamination ($3/21 \approx 14\%$) to the ULX catalog \citep{Walton2022}, consistent with their estimate of about 20\%. 
However, one should be cautious as we have selected ULXs with radio emission, and the fraction of quasars with detectable radio emission is in general much higher than that of ULXs. 
An unbiased estimate can be obtained only if this factor can be corrected.

\subsection{Radio emission due to star formation}

The radio counterparts of five ULXs appear to be spatially extended, including NGC 2207 ULX3, NGC 2903 ULX3, NGC 5457 ULX4, NGC 6926 ULX1, and PGC 90367 ULX1, whose integrated flux densities are 5.8, 7.9, 1.5, 12.0, and 2.3~mJy, respectively, in the same order at 3~GHz measured with VLASS, which has a beam size smaller than RACS.
They share a common property that the ULX resides in a star forming (\HII) region seen in optical.
Radio emission from \HII\ regions is composed of two components, thermal free-free emission which generally has a flat spectrum and non-thermal synchrotron radiation which has a negative slope \citep[typically $\alpha = -0.8$;][]{Condon1992, Murphy2011}.  
% \st{The thermal component usually dominates at frequencies above 1~GHz where the non-thermal component rapidly decays. }

For two of them, the star formation rate of the corresponding \HII\ region can be found in the literature, $1.6 \, M_\sun$~yr$^{-1}$ for NGC 2207 ULX3 \citep{Elmegreen2016, Elmegreen2017} and $0.126 \, M_\sun$~yr$^{-1}$ for NGC 2903 ULX3 \citep{Popping2010}. 
Given Eq.~(17) in \citet{Murphy2011} and the measured spectral index, the thermal plus non-thermal radio flux density due to star formation is expected to be 5.5 and 4.1 mJy at 3 GHz, suggesting that most or part of the observed radio emission could be due to star forming activities. 
% \st{However, we note that the two objects show a steep spectrum at frequencies up to 3 or 9~GHz} \red{\st{(no deviation from a single power-law in all bands)}}\st{, instead of a flat spectrum as expected for thermal emission above 1~GHz, casting doubt on the star forming scenario. }
Their spectral indices are generally consistent with that of synchrotron radiation found in star forming regions \citep{Tabatabaei2017, Galvin2018}, although $\alpha \approx -2$ seen in NGC 2903 ULX3 is steeper than typical.
Future high resolution radio observations are needed to investigate the nature of these sources. 
The X-ray emission with such a high luminosity is most likely coming from accreting compact objects in the \HII\ region. 
Future long-term monitoring is needed to distinguish whether there is a single ULX (likely the cases of NGC 2207 ULX3 and NGC 5457 ULX4) or multiple X-ray binaries (no variability is expected with a total number of $\sim$10--100 standard X-ray binaries to account for the observed luminosity). 

\subsection{Candidates for accreting compact objects in the host galaxy}

For the rest thirteen objects with a compact radio counterpart, 
eight of them (ESO 352-41 ULX1, NGC 925 ULX1, ESO 171-5 ULX1, NGC 6027 ULX1, IC 4587 ULX1, UGC 10143 ULX1, NGC 6946 ULX1 and NGC 7059 ULX1) show a steep spectrum ($\alpha \leq -0.5$), 
two of them (PGC 146838 ULX1 and NGC 5018 ULX6) display a flat or inverted spectrum ($\alpha \gtrsim 0$), and the other three sources do not have multiple frequency bands available for a slope measurement. 

Assuming that they reside in the apparent host galaxy, we discuss the nature of radio emission in two possible scenarios: jet/wind powered nebula due to supercritical accretion (SS~433/W50 like objects), or compact radio jets from accreting black holes.
Of course, it is also possible that the radio emission is unrelated to the X-ray source or accretion mechanism, but due to a compact star forming region as discussed above; this can be identified with future optical spectroscopy.

Radio emission from shock-ionized nebulae due to jets/winds is mainly optically thin synchrotron radiation with a steep spectrum. 
If SS 433 at a distance of 5 kpc \citep{Su2018} were observed in a nearby galaxy, it may appear as a small-scale or compact radio source, e.g., with an extension of 3\farcs6 at 10~Mpc or 0\farcs36 at 100 Mpc, and have a relatively weak flux density, e.g., 18~$\mu$Jy at 10~Mpc or 0.18~$\mu$Jy at 100~Mpc.
Of course, the size and flux depend on the jet/wind mechanical power.
In our sample, NGC 6946 ULX1 is a well studied, prototypical ULX and could be an interesting analog.  
However, its surrounding optical nebula MF16 is mainly powered by photoionization although shock-ionization is also important \citep{Abolmasov2007}.  
Thus, here we estimate the contribution of free-free emission to the observed radio flux based on the observed \hbeta\ luminosity. 
Adopting the relation between the \hbeta\ and radio emissivity of star forming regions \citep{Caplan1986}, we estimate that the radio flux density due to free-free emission is 0.08--0.12~mJy at 3~GHz given an \hbeta\ luminosity of $1.8 \times 10^{38}$~\ergs\ (translated to a distance of 7.7~Mpc assumed in this work) and a temperature in the range of $(0.9 - 2) \times 10^4$~K \citep{Abolmasov2008MF16}. 
Compared with the measurement of 0.76~mJy at 3~GHz, this suggests that only a small fraction of the radio emission can be attributed to free-free emission. 
It is worth noting that NGC 925 ULX1 is associated with an optical nebula (see \S~\ref{sec:ngc925}), further strengthening its analogy with NGC 6946 ULX1.
Therefore, the sources with a steep radio spectrum are good candidates for SS 433/W50 like objects in external galaxies and deserve in-depth multiwavelength studies in the future. 

Those with a flat or inverted spectrum resemble radio emission from a self-absorbed compact jet \citep{Blandford1979}.
Accreting black holes with radio emission in such a state are supposed to follow the fundamental plane of black hole activity \citep{Falcke2004, Corbel2013}.
Following \citet{Merloni2003}, we estimated a black hole mass as $1.3 \times 10^8$~$M_\sun$ and $4.0 \times 10^8$~$M_\sun$, respectively, for PGC 146838 ULX1 ($\alpha = -0.03\pm0.08$) and NGC 5018 ULX6 ($\alpha = 0.70\pm0.02$). 
Thus, they are all typical supermassive black holes in this scenario. 
We note that the X-ray luminosities listed in Table~\ref{tab:sample} are peak luminosities rather than the values measured simultaneously with the radio and would lead to an underestimate of black hole masses to some extent. 
Black holes of such high masses are not expected to appear at an off-nuclear position over the cosmological timescale, and would have spiraled into the center of galaxy due to dynamical friction \citep{Tremaine1975}. 
Thus, finding an off-nuclear supermassive black hole is of great interest and may shed light on the evolution history of the host galaxy, e.g., they could be tidally stripped nuclei residing in a star cluster \citep{Seth2014}.
We point out that these two objects as well as other objects discussed in this subsection all have an X-ray-to-optical flux ratio consistent with the range of AGNs \citep{Maccacaro1988}.
Optical spectroscopy can test if they are indeed AGNs via emission line properties (e.g., the presence of broad lines) and the BPT (Baldwin–Phillips–Terlevich) diagram \citep{Baldwin1981, Kewley2019, Law2021}. 

%% ESO 352-41 M_V<-11.3
%% PCG 146838 ULX2 M_V = -17

For the eight objects with a steep spectrum, we tend not to estimate their black hole mass using the fundamental plane. 
First, their radio emission may have an origin different from compact radio jets as discussed above, like the cases of NGC 925 ULX1 and NGC 6946 ULX1 where the radio emission is likely due to optically thin synchrotron radiation from the nebula. 
Second, these sources may be similar to compact steep-spectrum or peaked-spectrum radio sources \citep{Odea2021}, for which the fundamental plane is found to have a different relation \citep{Fan2016}.
We found that the inferred black hole masses of our sources using the relation in \citet{Fan2016} are 4--6 orders of magnitude lower than those using the relation in \citet{Merloni2003}. 
This is probably because the sample in \citet{Fan2016} have an X-ray luminosity mostly higher than $10^{42}$~\ergs, and the application of their relation is  a significant extrapolation in our cases. 

We note that both NGC 5018 ULX6 ($\alpha = 0.70\pm0.02$) and NGC 7059 ULX1 ($\alpha = -0.61\pm0.02$) displayed significant radio variability on a timescale as short as one day. 
This suggests that the radio emission cannot arise from a large structure, but constrained in a region smaller than $10^{15}$~cm. 

We also compared our sources with Fanaroff–Riley (FR) type 0 and type I radio galaxies in the plane of X-ray luminosity vs.\ radio luminosity \citep{Torresi2018}. 
There is only one object, IC 4587 ULX-1, well consistent with the FR0/FRI distribution in that plane.
All other objects display a radio luminosity generally below the luminosity range of the sample in \citet{Torresi2018} and have an X-ray-to-radio luminosity ratio on average higher the FR0/FRI galaxies.

HLX-1 has an optical magnitude $V = 24.5$ \citep{Soria2010HLX1}. 
Its X-ray peak luminosity is about $10^{42}$~\ergs\ \citep{Farrell2009}, with an X-ray-to-optical flux ratio of about 3.2, higher than that of any sources in this work. 
If HLX-1 appears at a distance of 30 Mpc or closer, one expects $V \leq 22$, detectable with the optical surveys used in this work.
Therefore, this work does not find any IMBH candidates typical of HLX-1 as we preferentially select radio-loud accretion systems. 
HLX-1 was found to display radio emission around 20 $\mu$Jy in its hard state \citep{Cseh2015}.
This approaches the detection limit of the two surveys if the source is located within 30~Mpc. 
It is interesting to note that the mass of HLX-1 estimated from the fundamental plane is about $3 \times 10^6$~$M_\sun$ using the relation in \citet{MillerJones2012} or $7 \times 10^5$~$M_\sun$ using that in \citet{Merloni2003}. 
These are both higher than the mass estimated from X-ray luminosities and temperatures \citep{Servillat2011, Soria2017}, typically on the order of $10^4$~$M_\sun$. 
Therefore, one should be cautious if the fundamental plane can be applied to these cases or if there is Doppler boosting in radio emission.

\section{Summary}
\label{sec:summary}

In this work, we found 21 ULXs with a radio counterpart with RACS and/or VLASS surveys:

\begin{itemize}\itemsep0em
    \item three are associated double lobed radio structures and may be quasars;
    \item five are associated with extended radio structure, and also spatially coincident with star forming regions in optical; their radio emission may be attributed to star forming activities but their steep spectrum indicates caveats; two of them show X-ray variability; the X-ray emissions are likely from ULXs in star forming regions; 
    \item thirteen are associated with an unresolved radio source:
    \begin{itemize}\itemsep0em
        \item eight have a steep radio spectrum, including one prototypical ULX (NGC 6946 ULX1), being good candidates for SS 433/W50 like objects;
        \item two have a flat or inverted spectrum and could be black holes with a compact jet; they are supermassive black holes based on the fundamental plane;
        \item significant radio variability on timescales as short as one day is detected in two sources (one has an inverted spectrum, and the other has a steep spectrum and is associated with gamma-ray emission); 
    \end{itemize}
    \item a redshift determination of these sources is of key importance.
\end{itemize}

\begin{acknowledgments}
We thank the anonymous referee for useful comments. HF acknowledges funding support from the National Natural Science Foundation of China under the grant 12025301, and the Strategic Priority Research Program of the Chinese Academy of Sciences.
\end{acknowledgments}

%%%%%%%%%%%%%%%%%%%%%%%%%%%%%%%%%%%%
\bibliographystyle{aasjournal}
\bibliography{refs}{}
%%%%%%%%%%%%%%%%%%%%%%%%%%%%%%%%%%%%
\end{document}